\documentclass[11pt,a4paper]{article}
\usepackage[utf8]{inputenc}
\usepackage{amsmath}
\usepackage{amsfonts}
\usepackage{amssymb}
\usepackage{amsthm} % added
\usepackage{graphicx}
\usepackage{subcaption} % for figure with subfigures.
\usepackage{tikz} % for some tikz figues.
\usepackage{placeins} % to use \FloatBarrier
\usepackage[version=3]{mhchem} % added for some of the reaction network graphs
\usepackage[linesnumbered,lined,boxed,commentsnumbered]{algorithm2e} % added for a nice looking algorithm environment
\usepackage{hyperref} % added
\RequirePackage{doi} % added
\usepackage[margin=3cm]{geometry}
\usepackage[sort&compress,comma,square,numbers]{natbib} % for the references and citation

\theoremstyle{plain}
\newtheorem{thm}{Theorem}[section]

\newtheorem{lem}[thm]{Lemma}

\theoremstyle{definition}
\newtheorem{definition}[thm]{Definition}
\newtheorem{remark}[thm]{Remark}

\newcommand{\R}{\mathbb{R}}
\newcommand{\N}{{\mathbb N}}
\newcommand{\Z}{{\mathbb Z}}
\newcommand{\cN}{\mathcal{N}}
\newcommand{\cS}{\mathcal{S}}
\newcommand{\cR}{\mathcal{R}}
\def\Vol{{\rm Vol}}
\def\Image{{\rm Im}}
\DeclareFontFamily{U}{FdSymbolC}{}
\DeclareFontShape{U}{FdSymbolC}{m}{n}{<-> s * FdSymbolC-Book}{}
\DeclareSymbolFont{fdarrows}{U}{FdSymbolC}{m}{n}
\DeclareMathSymbol{\leftfootline}{\mathrel}{fdarrows}{"AC}
\DeclareMathSymbol{\rightfootline}{\mathrel}{fdarrows}{"AD}
\DeclareMathSymbol{\longleftfootline}{\mathrel}{fdarrows}{"C6}
\DeclareMathSymbol{\longrightfootline}{\mathrel}{fdarrows}{"C7}

\newcommand{\footremember}[2]{%
	\footnote{#2}
	\newcounter{#1}
	\setcounter{#1}{\value{footnote}}%
}
\newcommand{\footrecall}[1]{%
	\footnotemark[\value{#1}]%
} 

\author{AmirHosein Sadeghimanesh\footremember{this}{Research Centre for Computational Sciences and Mathematical Modelling, Coventry University, Coventry, United Kingdom.}\footremember{Contact-AS}{amirhossein.sadeghimanesh@coventry.ac.uk.} and Matthew England\footrecall{this} \footremember{Contact-ME}{matthew.england@coventry.ac.uk, \url{https://www.matthewengland.coventry.domains/}.}}

\title{Polynomial Superlevel Set Representation of the Multistationarity Region of Chemical Reaction Networks}

\begin{document}

\maketitle

\begin{abstract}
	In this paper we introduce a new representation for the multistationarity region of a reaction network, using polynomial superlevel sets. The advantages of using this polynomial superlevel set representation over the already existing representations (cylindrical algebraic decompositions, numeric sampling, rectangular divisions) is discussed, and algorithms to compute this new representation are provided. The results are given for the general mathematical formalism of a parametric system of equations and so may be applied to other application domains.
	
\end{abstract}

\section{Introduction}
\label{sec:Introduction}

Many problems in applied sciences can be modelled by a parametric polynomial system, and therefore to solve such problems we must be able to explore the properties of these systems.  In particular, we often seek to identify areas of the parameter space where a property holds.  The contribution of this paper is a new methodology for exploring these. 

\subsection{Motivation:  Multistationarity Regions of CRNs}

We are motivated by the problem of understanding the multistationarity behavior of a Chemical Reaction Network (CRN). In a CRN, variables represent the concentrations of the species.  These change as time passes and are studied as part of the field of dynamical systems.  This is of polynomial type when the kinetics is assumed to follow the mass action rules. The equilibria of such a dynamical system are therefore the solutions to a system of polynomial equations.   However, the coefficients of the terms on the polynomials may involve some parameters. These parameters are usually the rates under which a reaction occurs and the total amounts (thought of as a dependency on the initial concentration of the species). 

Both the variables and parameters can only attain non-negative real values. A network is called multistationary if there exists a choice of parameters for which the network has more than one equilibrium. There are already many algorithms developed for answering the binary question of whether a system can exhibit multistationarity \cite{Feinberg-Deficiency-0-1-theorems,Feinberg-Deficiency-1-algorithm,Feinberg-Deficiency-one,Higher_deficiency_algorithm,Feliu-Sign,CRNToolbox,AmirSecondPaper,CoNtRol,Joshi-2013}. The input of these algorithms is a reaction network and the output is the confirmation or rejection of the possibility of exhibiting multistationary behavior. 

In the case where multistationarity can exist it then becomes important to determine the parameters where the network has this behavior. There has been less progress in this direction in the literature to date: the present paper offers a promising new development for this problem. 

\subsection{Prior Work}

Reviewing the state of the art in the literature, we see that in one vein of work focused on specific reaction networks with success following heuristic or manual calculations to find a suitable parameter which may not work in generality \cite{England-Errami-et-all-2020,N-site_2N-1_Conradi}. In another vein of work the system of equations for finding equilibria are solved for many random points from the parameter space to approximate the region where the network is multistationary \cite{Nam-Gyori-Amethyst-Bates-Gunawardeena-2020,Pusnik-Mraz-Zimic-Moskon-2019,Shah-Sarkar-2011,Chau-Walter-Gerardin-Tang-Lim-2012}.

Recently in \cite{Feliu-Sadeghimanesh-2020} a new approach to get a description of the multistationarity region is proposed. In this method one does not need to solve the system of equations to count the number of equilibriums. Instead one computes an integral to get the expected number of equilibriums when the parameters are following a random distribution.  This method partitions the parameter region into subsets that are a Cartesian product of intervals, called hyperrectangles. By choosing the uniform distribution and computing the average number of equilibriums on these hyperrectangles, one can approximate the multistationarity region as a union of sub-hyperrectangles. Which efficient and widely applicable, this list of hyperrectangles does not allow the reader much information or intuition about the geometry of this region, such as connectedness or convexity.

\subsection{Contribution}

In this work, we propose using polynomial superlevel sets to approximate the union of the hyperrectangles from \cite{Feliu-Sadeghimanesh-2020} as a set that can be described by one polynomial. Polynomial superlevel sets are already employed to approximate semi-algebraic sets and have been used in control and robust filtering contexts, see \cite{Dabbene-Henrion-Lagoa-2017,Dabbene-Henrion-2013}. The polynomial superlevel set representation we propose is a more compact representation of the region compared to a list of many hyperrectangles each described as a Cartesian product of intervals. Further, to check if a point belongs to the region given in this representation one can easily just evaluate the polynomial in this point. Further benefits of the polynomial superlevel set description of the region will be explored in Section \ref{sec:PSS}.

\subsection{Organization of the paper}\label{sec:Organization_of_the_paper}

The organization of this paper is as follows. The mathematical framework of reaction networks and the definition of the multistationarity region is given in Section \ref{sec:Multistationarity-Region}, while Section \ref{sec:Parametric_system_of_equations} contains the notations regarding parametric functions and definitions of the sampling and the rectangular representations of the multistationarity region from \cite[Section 2.4]{AmirPhDThesis}. 

In Section \ref{sec:PSS} we define polynomial superlevel sets formally and describe how one can algorithmically find a polynomial superlevel set representation of a set using the sampling and the rectangular representations. We demonstrate how to use it to find the polynomial superlevel set representation of the multistationarity region of a reaction network. Finally in Section \ref{sec:Bisect search} we discuss methods that sometimes can speed up computation of the polynomial superlevel set representation by the help of bisecting algorithms and where possible, algorithms for computing the expected number of solutions independently of solving the system itself.

\subsection{Notations}
\label{sec:Notations}

The cardinality of a set $A$ is denoted by $\#(A)$. Let $x\in\Z$ and
$n\in\Z\setminus\{0\}$. In this paper we define $x$ modulo $n$ to be $n$ instead of $0$ whenever $x$ is a multiple of $n$. For a function $f\colon A_1\rightarrow A_2$ and a point $u\in A_2$, the level set of $f$ is denoted by $L_u(f)$ and is defined as $\{x\in A_1\mid f(x)=u\}$. For two points $a=(a_1,\dots,a_n)$ and $b=(b_1,\dots,b_n)$ in $\R^n$, the notation $[a,b]$ is used to show the hyperractangle $\prod_{i=1}^n[a_i,b_i]$. For a subset $S$ of a hyperrectangle $B\subseteq\R^n$, let $\overline{\Vol}(S)$ denote the normalized volume of $S$ with respect to $B$, i.e.
\[\overline{\Vol}(S)=\frac{\Vol(S)}{\Vol(B)}.\]
When a random vector $X=(X_1,\dots,X_n)$ is distributed by a uniform distribution on a set $S\subseteq\R^n$, we write $X\sim U(S)$. If $X$ is distributed by a normal distribution with mean $\mu\in\R^n$ and variance $\sigma^2\in\R_{>0}$, then we write $X\sim N(\mu,\sigma^2)$ and we mean that $X_1,\dots,X_n$ are identically and independently distributed by $N(\mu_i,\sigma^2)$. The expectation of $g(X)$ when $X$ is distributed by a probability distribution $q$ is denoted by $\mathbb{E}\big(g(X)\mid X\sim q\big)$.
\medskip

\subsection{Computer information}

All computations for the examples of this paper were done on a computer with the following information.   
\texttt{Processor:} \texttt{Intel(R)} \texttt{Core(TM)} \texttt{i7-10850H} \texttt{CPU} \texttt{@2.70GHz} \texttt{2.71 GHz}.  
\texttt{Installed} \texttt{memory} \texttt{(RAM):} \texttt{64.0 GB} \texttt{(63.6} \texttt{GB} \texttt{usable)}.  
\texttt{System} \texttt{type:} \texttt{64-bit} \texttt{Operating} \texttt{System,} \texttt{x64-based} \texttt{processor.}\\

The software and programming languages used for the computations reported in this paper had the following version numbers:
\texttt{Maple 2021}, \texttt{Matlab R2021a}, \texttt{YALMIP}, \texttt{SeDuMi 1.3}, \texttt{Julia 1.6.2}, \texttt{MCKR 1.0}.

\section{Multistationarity region of chemical reaction networks}
\label{sec:Multistationarity-Region}

In this section, we introduce the concepts of reaction network theory that are needed throughout the rest of the paper, with the help of a simple gene regulatory network example. 

One can think of a gene as a unit encoding information for the synthesis of a product such as a protein. First, a group of DNA binding proteins called transcription factors bind a region of the gene called promoter. Now an enzyme called RNA polymerase starts reading the gene and produces an RNA until it arrives in the terminator region of the gene. The process until here is called the transcription step. After transcription got completed, the resulting RNA leaves the nucleus (in eukaryotes) and reaches ribosomes. In ribosomes, the second step, called translation, gets started. Ribosomes assemble a protein from amino acids using the manual guide written in the RNA. A gene encoding of a protein recipe is said to be expressed when it gets transcribed to an RNA, and the RNA translated to the protein. A gene is not always expressed in a constant rate. There might be proteins that bind the transcription factors or the promoter region and, as a result, inhibit the RNA polymerase starting the transcription process.
On the other hand, there might be other proteins in which their binding to the transcription factors or the promoter region enhances the transcription.

Consider a simple example from \cite[Figure 2]{ThreeGenes-GeneRegulatory-Nature}, depicted here in Figure \ref{fig:3-gene-network}. There are three genes with proteins $A$, $B,$ and $C$ as their final products. Denote their concentrations at time $t$ by $[A](t)$, $[B](t)$ and $[C](t)$ respectively. The concentration of these proteins will not remain constant all the time, and we have an Ordinary Differential Equation (ODE) system describing the variation of the concentrations as time passes, see Figure \ref{fig:3-gene-ode}. Each protein is degraded with a first-order kinetics with the reaction rate constants $k_{A,d}$, $k_{B,d}$ and $k_{C,d}$ correspondingly. Protein $A$ activates the expression of the second gene with Michaelis-Menten kinetics with the maximum rate $k_{B,\max}$ and the Michaelis constant $k_{B,A}^{-1}$. The third gene gets activated by both proteins $A$ and $B$ together with the product of two Michaelis-Menten kinetics, with maximum rate $k_{C,\max}$ and Michaelis constants $k_{C,A}^{-1}$ and $k_{C,B}^{-1}$. The first gene gets expressed by the rate $k_{A,\max}$ in the absence of protein $C,$ and protein $C$ has an inhibitory effect on the expression of the first gene, captured by the denominator $(1+k_{A,C}[C](t))$ in the rate expression.

\begin{figure*}[ht]
	\centering
	\begin{subfigure}[b]{0.3\textwidth}
		\resizebox{\linewidth}{!}{
			\begin{minipage}{\linewidth}
				\begin{tikzpicture}[
					roundnode/.style={circle, draw=green!60, fill=green!5, very thick, minimum size=7mm},
					squarednode/.style={rectangle, draw=red!60, fill=red!5, very thick, minimum size=5mm},
					]
					%Nodes
					\node[]      (gene1)    at (0,2)  {Gene 1};
					\node[]      (gene2)    at (2,2)  {Gene 2};
					\node[]      (gene3)    at (4,2)  {Gene 3};
					\node[]                 (joint1)   at (2,3)  {}; 
					\node[]                 (joint2)   at (3,1)  {};
					%\node[<options>](<name>) at (<coordinate>){<text>};
					%Arrows
					\draw[-|]  (gene3.north) .. controls +(up:7mm) and +(right:7mm) .. (joint1.east) .. controls +(left:7mm) and +(up:7mm) .. (gene1.north) ;
					%\draw[-|] (joint1.west) -- (gene1.north);
					\draw[->] (gene1.east) -- (gene2.west);
					\draw[-]  (gene2.south) .. controls +(down:7mm) and +(left:3mm) .. (joint2.west); 
					\draw[->]  (gene1.south) .. controls +(down:7mm) and +(left:14mm) .. (joint2.west) .. controls +(right:7mm) and +(down:7mm) .. (gene3);
				\end{tikzpicture}
			\end{minipage}
		}
		\vspace{0.25cm}
		\caption{}
		\label{fig:3-gene-network}
	\end{subfigure}
	\hfill
	\begin{subfigure}[b]{0.62\textwidth}
		\centering
		\resizebox{\linewidth}{!}{
			\begin{minipage}{\linewidth}
				\begin{align*}
					\frac{d[A](t)}{dt} &= k_{A,\max}\cdot\dfrac{1}{1+k_{A,C}\cdot[C](t)}-k_{A,d}\cdot[A](t)\\
					\frac{d[B](t)}{dt} &= k_{B,\max}\cdot\dfrac{k_{B,A}\cdot[A](t)}{1+k_{B,A}\cdot[A](t)}-k_{B,d}\cdot[B](t)\\
					\frac{d[C](t)}{dt} &= k_{C,\max}\cdot\dfrac{k_{C,A}\cdot[A](t)}{1+k_{C,A}\cdot[A](t)}\cdot\dfrac{k_{C,B}\cdot[B](t)}{1+k_{C,B}\cdot[B](t)}-k_{C,d}\cdot[C](t)
				\end{align*}
			\end{minipage}
		}
		\caption{}
		\label{fig:3-gene-ode}
	\end{subfigure}
	\caption{A regulatory network of 3 genes \cite[Figure 2]{ThreeGenes-GeneRegulatory-Nature}. \\
		(a) This graph shows the relations between expressions of the genes. We denote by $\rightfootline$ an inhibitory relation and by $\rightarrow$ a positive relation. \\
		(b) The system of Ordinary differential equations for the network.}
\end{figure*}

A solution to the system $\frac{d[X_i](t)}{dt}=0$ (where the $X_i$s are $A$, $B$ and $C$) is called an equilibrium of the ODE system. Since the concentration of the proteins can only be non-negative real numbers, the complex or negative real solutions are not relevant. Sometimes we may only consider the positive solutions, for example, if a total consumption of a protein is not possible or of no interest. Therefore by steady states we mean positive solutions to the system of equations $\frac{d[X_i](t)}{dt}=0$. The equations in this system are called the steady state equations.

Now we are ready to define a reaction network formally. A reaction network, or a network for short, is an ordered pair, $\cN=(\cS,\cR)$ where $\cS$ and $\cR$ are two finite sets called the set of species and the set of reactions. In our example, $\cS=\{A,B,C\}$ and $\cR$ contains six reactions: three gene expressions and three protein degradations. To each network, an ODE is attached with concentration of the species as its variables and the constants of the reaction rate expressions as its parameters. In our example, we have such 3 variables and $10$ parameters. To fix the notation assume $S=\{X_1,\dots,X_n\}$ and that there are $r$ constants involved in the reaction rate expressions. Then we use $x_i$ instead of $[X_i](t)$ and $k_i$ for the $i$-th parameter. Denote by $f_{i,k}(x)$ the $i$-th steady state equation where $x=(x_1,\dots,x_n)$ and $k=(k_1,\dots,k_r)$.

A network with an inflow (injection) or an outflow (extraction or degradation) for at least one of its species is called an \emph{open network}. The network in Figure \ref{fig:3-gene-network} is an open network because of the presence of the degradation reactions. A network can also be fully or partially conserved. 

Consider the simple single reaction network depicted in Figure \ref{fig:H2O2-network}. The system of its ODE equations is given in Figure~\ref{fig:H2O2-ODE}. Because $\dot{x_1}+2\dot{x_3}=0$, the linear combination $x_1+2x_3$ should be constant with respect to the time. Therefore there exists a positive constant $T_1$ such that the relation $x_1+2x_3=T_1$ holds. Similarly there exist two other positive constants $T_2$ and $T_3$ such that $x_1+2x_4=T_2$ and $x_2+2x_3=T_3$. The values of $T_1$, $T_2$ and $T_3$ can be determined by the initial conditions of the ODE system. These linear invariants imply that three of the steady state equations are linearly redundant and can be replaced by these three linear invariants which are called \emph{conservation laws} in CRN theory. The linear subspace determined by the conservation laws is called the \emph{stoichiometric compatibility class}. For a more detailed definition of conservation laws see Definition 1 in \cite[Chapter 2]{AmirPhDThesis}. One should note that the trajectories of the ODE system are confined to stoichiometric compatibility classes. In this case, one only cares about the steady states in one stoichiometric compatibility class.

\begin{figure*}[ht]
	\centering
	\begin{subfigure}[b]{0.4\textwidth}
		\resizebox{\linewidth}{!}{
			\begin{minipage}{\linewidth}
				\ce{2O2^- + 2H^+ ->[k] O2 + H2O2}
			\end{minipage}
		}
		\vspace{0.25cm}
		\caption{}
		\label{fig:H2O2-network}
	\end{subfigure}
	\hfill
	\begin{subfigure}[b]{0.52\textwidth}
		\centering
		\resizebox{\linewidth}{!}{
			\begin{minipage}{\linewidth}
				\begin{align*}
					\frac{dx_1}{dt}=-2kx_1^2x_2^2, &\quad \frac{dx_3}{dt}=kx_1^2x_2^2 \\
					\frac{dx_2}{dt}=-2kx_1^2x_2^2, &\quad \frac{dx_4}{dt}=kx_1^2x_2^2	
				\end{align*}
			\end{minipage}
		}
		\caption{}
		\label{fig:H2O2-ODE}
	\end{subfigure}
	\caption{A simple example of a closed network consisting of one reaction. \\
		(a) Two molecules of superoxide and two hydron atoms react to each other and produce one molecule of dioxygen and a molecule of hydrogen perixide. The reaction rate here follows the mass-action kinetics with the reaction rate constant $k$. \\
		(b) The system of ODE equations of the network. The concentrations of \ce{O2-}, \ce{H+}, \ce{O2} and \ce{H2O2} are denoted by $x_1$, $x_2$, $x_3$ and $x_4$ respectively.}
\end{figure*}

Now we are ready to define the main concept of interest, multistationarity.

\begin{definition}\label{def:Multistationarity}
	Consider a network with $n$ species and replace redundant steady state equations by conservation laws if there exist any. Let $k$ stands for the vector of constants of both the reaction rates and conservation laws and be of the size $r$. A network is called \emph{multistationary} over $B\subseteq\R^r$ if there exists a $k\in B$ such that $f_k(x)=0$ has more than one solution in $\R_{>0}^n$.
\end{definition}

\begin{remark}\label{remark:conservation-laws-and-kinetics-choice}
	\hfill
	\begin{itemize}
		\item[i)] One may also consider non-linear invariants such as first integrals as defined in \cite[Definition 11]{FirstIntegralsDefinition}.
		\item[ii)] Note that we are not concerned with the choice of the kinetics such as mass-action, Michaelis-Menten, Hill function, power-law kinetics and S-systems, or the form of the steady state equations such as polynomial or  rational functions. Therefore the results of this paper will remain valid and practical for a general reaction network.
	\end{itemize}
\end{remark}

From here on the word parameters also includes the constants of the conservation laws in addition to the reaction rate constants. To answer the question of whether a network is multistationary or not one can use one of many algorithms available in the literature, see \cite{Feinberg-Deficiency-0-1-theorems,Feinberg-Deficiency-1-algorithm,Feinberg-Deficiency-one,Higher_deficiency_algorithm,Feliu-Sign,AmirSecondPaper,Joshi-2013} for a few examples. However, to partition the parameter space into two subsets, one consisting of the choices of parameters for which $f_k(x)$ has more than one solution and the other comprising those parameter choices for which $f_k(x)$ has at most one solution, is a more laborious task which we tackle in this paper.

\begin{definition}\label{def:Multistationarity-region}
	Consider a reaction network with the setting and notation of Definition \ref{def:Multistationarity}. The set 
	\[
	\{k\in B\mid \#\big(f_k^{-1}(0)\cap \R_{>0}^n\big)\geq 2\} 
	\]
	is called the multistationarity region of the network.
\end{definition}

The region $B$ in Definition \ref{def:Multistationarity-region} represents the regions of scientific interest.  It will usually be a hyperrectangle made by the inequality restrictions of the form $k_{i,\min}<k_i<k_{i,\max}$ for the parameters. This is because, for example the rate of expression of a gene can not be any arbitrary positive number but must be limited; or the constant of conservation laws may be limited from above due to the limitation of the materials in the lab.

\section{Prior state-of-the-art for parametric systems of equations}
\label{sec:Parametric_system_of_equations}

Let $f_k\colon\R^n\rightarrow\R^m$ be a parametric function with $B\subseteq\R^r$ as its parameter region and $u$ a point in $\R^m$. For each choice of the parameters $k^\star\in B$, the system $f_{k^\star}(x)=u$ is a non-parametric system of equations. One can solve this system and look at the cardinality of the solution set. For different choices of $k^\star$, this number can be different. Therefore we define a new function $\Phi_f^u\colon B\rightarrow \Z_{\geq 0}\cup\{\infty\}$ sending $k\in B$ to $\#\big(L_u(f)\big)$, i.e. the size of the level set of $f_k$ (the set of points in $\R^n$ which $f_k$ maps to $u$). 
Now one can partition  $B$ into the union of level sets of the map $\Phi_f^u$. 
For a general form of $f_k(x)$, finding $L_i(\Phi_{f}^u)$ is a hard question.

\subsection{CAD with respect to discriminant variety}

In the case where $f_k(x)\in\big(\R(k)[x]\big)^m$ and $A$ and $B$ are semi-algebraic sets there are a variety of tools which can be employed, see for example \cite{BPR06}.  In the literature, the approach used most commonly (e.g. \cite{England-Errami-et-all-2020,Lichtblau-2021,Rost-AmirHosein-2021}) is a Cylindrical Algebraic Decomposition (CAD) computed with respect to the discriminant variety.  For a full description of this technique we refer the reader to \cite{Lazard-Rouillier-SolvingParametricPolynomialSystemsFabrice,Moroz-PhD-thesis} or the short sketch of the main idea in \cite[section 3]{Rost-AmirHosein-2021}. Briefly: the discriminant variety of the system $f_k(x)$ with the domain and codomain restrictions on the semi-algebraic sets $A$ and $B$ is the solution set to a new set of (non-parametric) polynomial equations with $k$ as its indeterminants. This new set of polynomials can be computed algorithmically for example using Gr\"obner bases and elimination theory. Then CAD with respect to the discriminant variety decomposes $B$ into a finite number of connected semi-algebraic sets called cells. Each cell has intersection with only one $L_i(\Phi_f^u)$ and therefore $L_i(\Phi_f^u)$ can be expressed as union of a finite number of cells with an exact description of their boundaries. 

As we see later in this section, in many cases one is only interested in open cells (i.e. only those cells which have full dimension \cite{WBDE14}).  A Maple package, \texttt{RootFinding[Parametric]} has implemented an algorithm to compute the open CAD with respect to the discriminant variety of a system of parametric polynomial equations and inequalities \cite{Gerhard-Jeffrey-Moroz-2010}. From here on in this paper by \emph{CAD} we mean such an open CAD with respect to the discriminant variety.

Both the computation of the discriminant variety and the subsequent decomposition involve the use of algorithms with doubly exponential complexity which can cause problems.  The number of cells in the decomposition will grow doubly exponentially in the number of parameters of $f_k(x)$ \cite{Bradford-Davenport-England-McCallum-Wilson-2016}; and even computation of the discriminant variety itself before any decomposition can be infeasible for moderate examples, see e.g. \cite{Rost-AmirHosein-2021}.  This makes CAD impractical for studying parametric systems of polynomial equations with more than a few variables and parameters.

\subsection{Approximation by sampling}
\label{ssec:sampling}

Another approach adopted by scientists is to solve the system $f_k(x)=u$ for many different choices of $k\in B$ \cite{Nam-Gyori-Amethyst-Bates-Gunawardeena-2020,Pusnik-Mraz-Zimic-Moskon-2019,Shah-Sarkar-2011,Chau-Walter-Gerardin-Tang-Lim-2012}.

Mathematically speaking, this means that $B$ is replaced by a finite set. Then each $L_i(\Phi_f^u)$ is expressed as a subset of this finite set. This approach hereafter is referred as the \emph{sampling representation} approach. In contrast with the CAD approach which provides an exact description of $L_i(\Phi_f^u)$, the sampling representation approach provides an approximation. Note that there are different ways to choose the sample parameter points for the sampling representation. One way is to arrange all points equally distanced like a grid, and another way is to randomly sample from a distribution such as the uniform distribution on $B$, which is the one used in this paper. For an example of a case where a sampling representation with grid-like parameter sampling is used see \cite[Figures $7-12$]{England-Errami-et-all-2020}.

Since we are motivated from the application, we should note that in a lab, it is usually not possible to design the experiment so that the parameter values are exactly the numbers that we decide. Therefore when the experiment is designed to have $k=k^\star$, what happens is that $k$ is a point in a neighborhood of $k^\star$ and not necessarily $k^\star$ itself. This can happen for example because of errors coming from the measurement tools or the noise from the environment. In such cases picking a point close to the boundaries of $L_i(\Phi_f^u)$ could lead to a different result than what the experimentalist expects, if errors or noise push it over the boundary.

\subsection{Rectangular representation}

A different discretization can be done using a rectangular division of $B$. For example if $B$ is a hyperrectangle $[a,b]$ then a grid on $B$ is achieved by dividing $B$ along each axis to equal parts. Then for each sub-hyperrectangle of $B$ in this rectangular division we assign the average of the number of solutions of $f_k(x)=u$ for several choices of $k$ coming from the sub-hyperrectangle. This approach hereafter is referred as the \emph{rectangular representation} approach. See Figure \ref{fig:Gene_Regulatory_BMC_dimer_ThreeRepresentations} to compare the three approaches visually.

\subsection{Example}
\label{example:CAD_Finite_Grid}

Consider the gene regulatory network in 
\cite[Figure 3B]{CapacityOfMultistationaritySmallGeneRegulatory}, depicted here in Figure \ref{fig:Gene_Regulatory_BMC_dimer-network} with the ODE system in Figure \ref{fig:Gene_Regulatory_BMC_dimer-ODE}. This network has one conservation law, $x_1+x_4=k_8$. Therefore we consider the system of equations obtained by the first three steady state equations in the ODE system and the conservation law to study the multistationarity of this network. For illustration purpose we fix values of all parameters other than two so we can plot the multistationarity region in 2 dimensions. In \cite[Figure 4]{CapacityOfMultistationaritySmallGeneRegulatory} the reaction rate constants other than $k_3$ were fixed to the values listed below.
\begin{equation}\label{eq:parameters_values_for_Gene_Regulatory_BMC_dimer}
	k_1=2.81,\;k_2=1,\;k_4=0.98,\;k_5=2.76,\;k_6=1.55,\;k_7=46.9.
\end{equation}
We fix the values of all parameters other than $k_3$ and $k_8$ to these values also.

\begin{figure*}[ht]
	\centering
	\begin{subfigure}[b]{0.4\textwidth}
		\resizebox{\linewidth}{!}{
			\begin{minipage}{\linewidth}
				\[\begin{array}{c}
					X\ce{->[k_1]}X+P\\
					P\ce{->[k_2]}0\\
					2P\ce{<=>[k_3][k_4]}PP\\
					X+PP\ce{<=>[k_5][k_6]}XPP\\
					XPP\ce{->[k_7]}XPP+P
				\end{array}\]
			\end{minipage}
		}
		\vspace{0.25cm}
		\caption{}
		\label{fig:Gene_Regulatory_BMC_dimer-network}
	\end{subfigure}
	\hfill
	\begin{subfigure}[b]{0.52\textwidth}
		\centering
		\resizebox{\linewidth}{!}{
			\begin{minipage}{\linewidth}
				\begin{align*}
					\frac{dx_1}{dt} &= -k_5x_1x_3+k_6x_4,\\ 
					\frac{dx_2}{dt} &= k_1x_1-k_2x_2-2k_3x_2^2+2k_4x_3+k_7x_4,\\
					\frac{dx_3}{dt} &= k_3x_2^2-k_4x_3-k_5x_1x_3+k_6x_4,\\ \frac{dx_4}{dt} &= k_5x_1x_3-k_6x_4	
				\end{align*}
			\end{minipage}
		}
		\caption{}
		\label{fig:Gene_Regulatory_BMC_dimer-ODE}
	\end{subfigure}
	\caption{A bistable autoregulatory motif presented in \cite[Figure 3B]{CapacityOfMultistationaritySmallGeneRegulatory}. \\
		(a) $X$ is a gene, $P$ is a protein that can form a dimer $PP$ and then binding to $X$. The gene $X$ will get expressed and produce $P$ in both forms $X$ and $XPP$. Finally there is a degradation of $P$. \\
		(b) The ODE system of the gene regulatory network in part (a). The variables $x_1$, $x_2$, $x_3$ and $x_4$ are standing for the concentration of the species $X$, $P$, $PP$ and $XPP$ respectively.}
\end{figure*}

\begin{figure*}[ht]
	\centering
	\begin{subfigure}[b]{0.31\textwidth}
		\includegraphics[width=\linewidth]{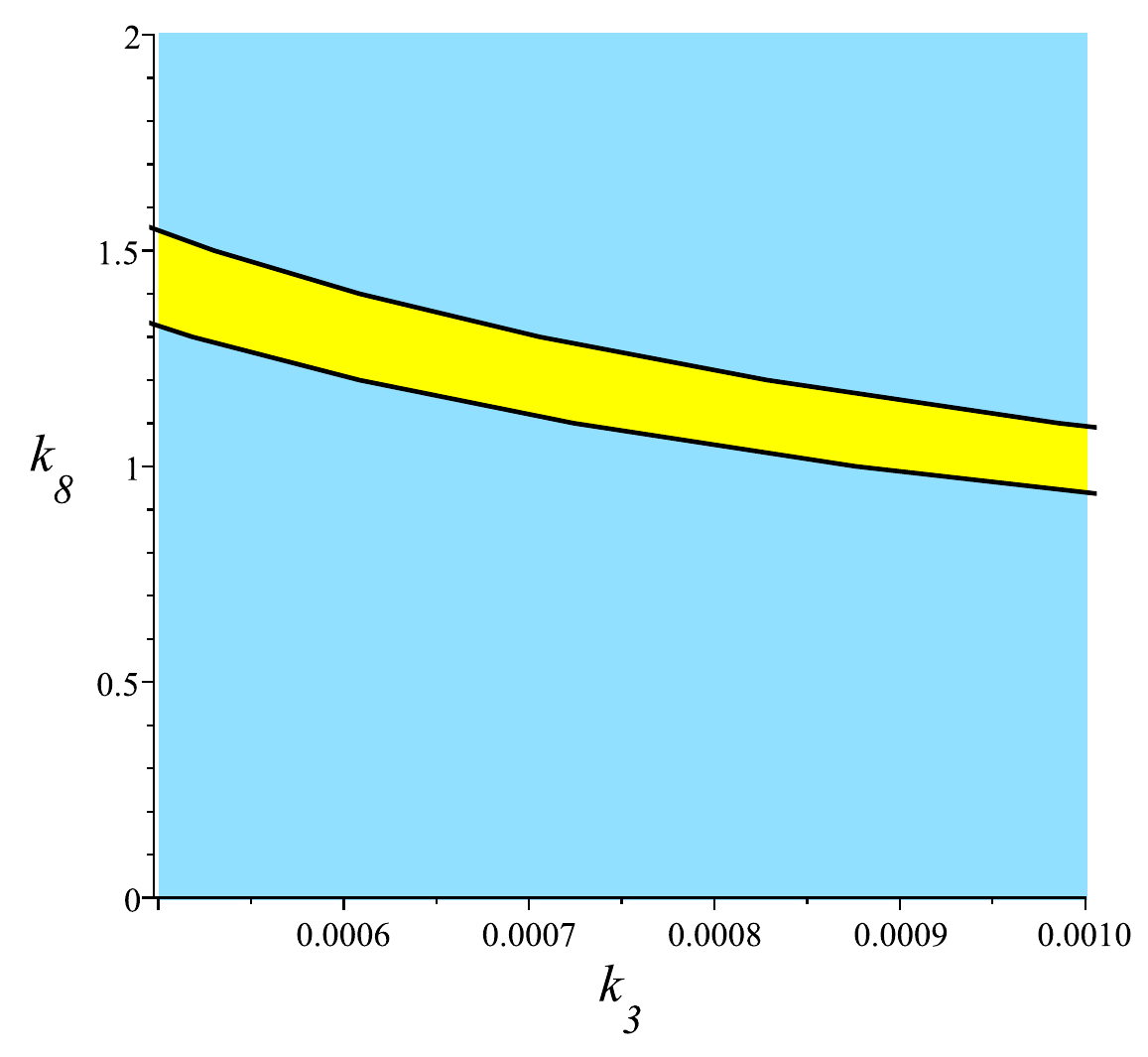}
		\caption{}
		\label{fig:Gene_Regulatory_BMC_dimer-CAD}
	\end{subfigure}
	\hfill
	\begin{subfigure}[b]{0.31\textwidth}
		\includegraphics[width=\linewidth]{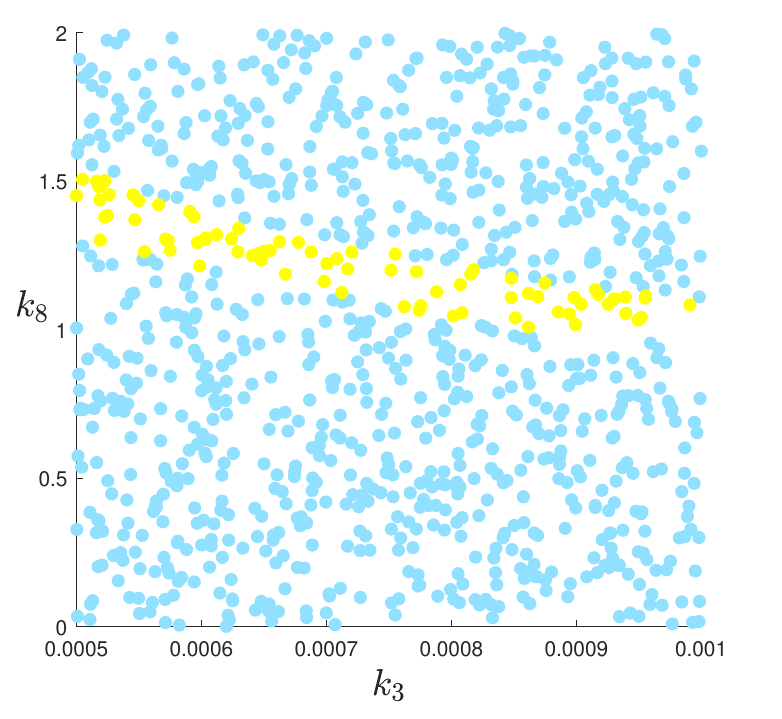}
		\caption{}
		\label{fig:Gene_Regulatory_BMC_dimer_FiniteRepresentation}
	\end{subfigure}
	\hfill
	\begin{subfigure}[b]{0.33\textwidth}
		\includegraphics[width=\linewidth]{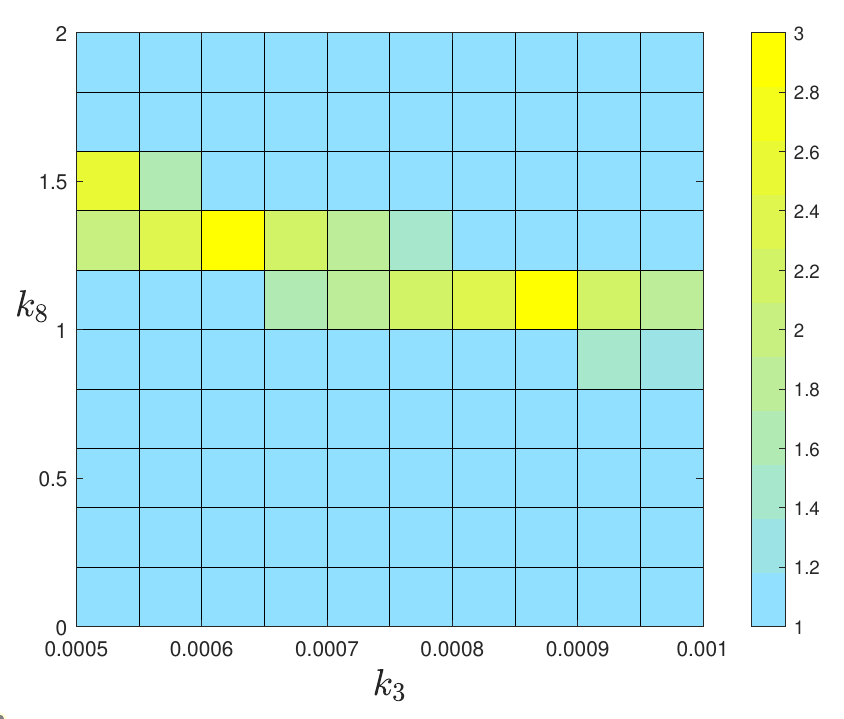}
		\caption{}
		\label{fig:Gene_Regulatory_BMC_dimer_GridRepresentation}
	\end{subfigure}
	\caption{Three representations of the multistationarity region of the network in Figure \ref{fig:Gene_Regulatory_BMC_dimer-network} after fixing all parameter values other than $k_3$ and $k_8$ to the values in \eqref{eq:parameters_values_for_Gene_Regulatory_BMC_dimer}. \\
		(a) CAD gives the exact boundary of $L_1(\Phi_f^0)$ and $L_3(\Phi_f^0)$.  The first one is colored by sky blue and the later one with yellow. \\
		(b) A sampling representation of the parameter region $B=[(0.0005,0),(0.001,2)]$ by 1000 random points sampled from uniform distribution on $B$. 78 of these points belong to $L_3(\Phi_f^0)$ and are colored yellow. The other 922 points belong to $L_1(\Phi_f^0)$ and are colored by sky blue. \\
		(c) A rectangular representation of $B$. Each subrectangle is colored with respect to the average number of solutions for 10 random points sampled from the uniform distribution on the subrectangle. The color bar of the figure is in the right side.}
	\label{fig:Gene_Regulatory_BMC_dimer_ThreeRepresentations}
\end{figure*}

Let $B$ be the rectangle made by the constraints $0.0005<k_3<0.001$ and $0<k_8<2$. Using the \texttt{RootFinding[Parametric]} package of Maple we get the exact description of the multistationarity region of the network, in 0.12 seconds, depicted in Figure \ref{fig:Gene_Regulatory_BMC_dimer-CAD}.

A sampling representation of the multistationarity region is found by solving the system of the equations for 1000 points $(k_3,k_8)$ sampled from the uniform distribution on $[(0.0005,0),(0.001,2)]$. We used the \texttt{vpasolve} command from Matlab to solve the system numerically. The Matlab code to generate this sampling representation took 154 seconds to run, with the output visualised in Figure \ref{fig:Gene_Regulatory_BMC_dimer_FiniteRepresentation}.

A rectangular representation is given by dividing $[(0.0005,0),(0.001,2)]$ to 100 equal sub-rectangles and then solving the system for 10 points $(k_3,k_8)$ sampled from the uniform distribution on each sub-rectangles. The sub-rectangles are colored with respect to the average number of solutions. This computation also was done by Matlab and took 166 seconds, with the output visualised in Figure \ref{fig:Gene_Regulatory_BMC_dimer_GridRepresentation}.

\section{Polynomial superlevel set representation}
\label{sec:PSS}

\subsection{Superlevel sets} 

\begin{definition}\label{def:PSS}
	Consider $f\colon\R^n\rightarrow\R$, an arbitrary function. For a given $u\in\R$ a \emph{superlevel set} of $f$ is the set of the form
	\[U_u(f)=\{x\in\R^n\mid f(x)\geq u\}.\]
	When $u=1$ we drop the index and write only $U(f)$. Naturally, a \emph{polynomial superlevel set} is a superlevel set of a polynomial.
\end{definition}

Polynomial sublevel sets are defined similarly as in Definition \ref{def:PSS} with the only difference the direction of the inequality. However, in this paper, we only focus on superlevel sets. For $d\in\Z_{\geq 0}$ let $P_d$ denote the set of polynomials of total degree at most $d$. A sum of squares (SOS) polynomial of degree $2d$ is a polynomial $p\in P_{2d}$ such that there exist $p_1,\dots,p_m\in P_d$ so that $p=\sum_{i=1}^m p_i^2$. We denote the set of SOS polynomials of degree at most $2d$ by $\Sigma_{2d}$.

\begin{thm}[{\cite[Theorem 2]{Dabbene-Henrion-Lagoa-2017}}]\label{thm:PSS}
	Let $B\subseteq\R^n$ be a compact set and $K$ a closed subset of $B$. For $d\in\N$ define
	\[S_d=\{p\in P_d\mid p\geq 0\text{ on }B,\;p\geq 1\text{ on }K\}.\]
	Then there exists a polynomial $p_d\in S_d$ such that
	\[
	\int_Bp_d(x)dx=\inf\Big\{\int_Bp(x)dx\mid p\in S_d\Big\}.
	\]
	Furthermore $\lim_{d\rightarrow\infty}\Vol(U(p_d)-K)=0$.
\end{thm}

Given a pair $(B,K)$ where $B\subseteq\R^n$ is a compact set and $K\subseteq B$ a closed set, and given $d\in\N$; we call the polynomial superlevel set $U(p)$ (with $p$ being the polynomial $p_d\in S_d$ found in Theorem \ref{thm:PSS}) the \emph{PSS representation} of $K\subseteq B$ of degree $d$. When $K$ is a semi-algebraic set, one can find $p_d$ numerically using a minimization problem subject to some positivity constraints \cite[Equation 13]{Dabbene-Henrion-2013}. 

Let $B=[a_B,b_B]$ and $K_i=[a_{K_i},b_{K_i}]$, $i=1,\dots,m$ be some hyperrectangles in $\R^n$ such that $K:=\cup_{i=1}^mK_i\subseteq B$. By solving a similar optimization problem it is possible to find the PSS representation of $K\subseteq B$. Let $d\in\N$. The goal is to find the coefficients of a polynomial of degree $d$ such that $\int_Bp(x)dx$ becomes minimum subject to some conditions. Before presenting the constraints, let us look at the target function. A polynomial $p(x)$ of degree $d$ can be written as $\sum_{\alpha\in\N_d^n}c_\alpha x^\alpha$. Here $\N_d^n$ is the set of $\alpha=(\alpha_1,\dots,\alpha_n)\in \Z_{\geq 0}^n$ such that $\sum_{i=1}^n\alpha_i\leq d$. Now the integral can be simplified as below.
\begin{align*}
	\int_Bp(x)dx &= \int_B\Big(\sum_{\alpha\in\N_d^n}c_\alpha x^\alpha\Big)dx \\
	&= \sum_{\alpha\in\N_d^n}c_\alpha\int_Bx^\alpha dx \\
	&= \sum_{\alpha\in\N_d^n}\Big(\int_Bx^\alpha dx\Big)c_\alpha.
\end{align*}
Since $\int_Bx^\alpha dx$ are constant real numbers independent of the coefficients of the polynomial, the target function is a linear function on the coefficients of $p(x)$ which are the variables of the optimization problem. 

Now let us look at the constraints. First of all $p(x)$ has to be nonnegative on $B$. This can be enforced by letting
\begin{equation}
	\label{eq:constraints_nonnegative_on_hyperrectangle}
	p(x)-\sum_{j=1}^ns_{B,j}(x)\big(x_j-a_{B,j}\big)\big(b_{B,j}-x_j\big)\in\Sigma_{2r},\quad s_{B,j}\in\Sigma_{2r-2},j=1,\dots,n,
\end{equation}
where $r=\lfloor\frac{d}{2}\rfloor$ the largest integer less than or equal to $\frac{d}{2}$. Secondly we need $p(x)\geq 1$ on $K$ or in other words $p(x)-1\geq 0$ on $K$. This holds if and only if $p(x)-1\geq 0$ on each $K_i$. Therefore for every $i=1,\dots,m$ one more constraint of the shape \eqref{eq:constraints_nonnegative_on_hyperrectangle} has to be added:
\begin{equation*}
	p(x)-1-\sum_{j=1}^ns_{K_i,j}(x)\big(x_j-a_{K_i,j}\big)\big(b_{K_i,j}-x_j\big)\in\Sigma_{2r},
	\quad s_{K_i,j}\in\Sigma_{2r-2},j=1,\dots,n.
\end{equation*}

Recall Definition \ref{def:Multistationarity-region}: the multistationarity region of a network is in fact a superlevel set, $U_2(\Phi_f^0)$. The goal is to find a PSS representation of the set $U_2(\Phi_f^0)$. One way to accomplish this goal is to find a rectangular representation of the multistationarity region and then solve the above mentioned SOS optimization problem. The next example illustrates this idea.  To tackle it we use a Matlab toolbox called \texttt{YALMIP} \cite{YALMIP,How_To_solve_SOS} which can receive an SOS optimization problem, process it and use other solvers to solve it.  For the solver to be used by \texttt{YALMIP}, we chose \texttt{SeDuMi} \cite{SeDuMi}.

\subsection{Examples}
\label{example:Comparision-PSS-grid}

\begin{figure*}[ht]
	\centering
	\includegraphics[width=5cm]{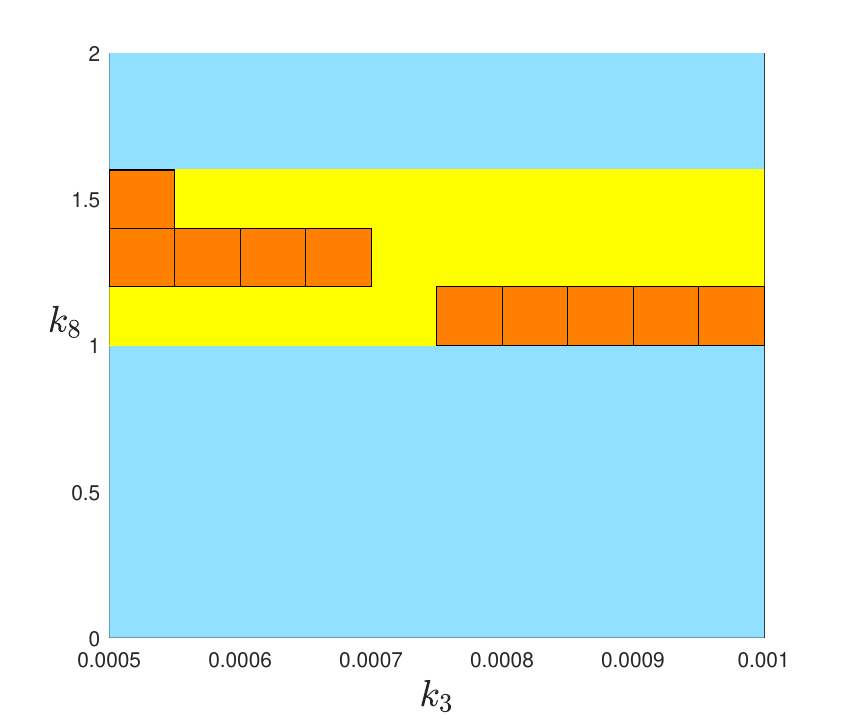}
	\caption{The PSS approximation of the multistationarity region for the network in Figure \ref{fig:Gene_Regulatory_BMC_dimer-network} of degree 2 obtained by the information of Figure \ref{fig:Gene_Regulatory_BMC_dimer_GridRepresentation}. The union of orange colored subrectangles is considered as the initial approximation of the multistationarity region obtained by the rectangular representation and chosen as the set $K$. The yellow colored area is the difference of $U(p)-K$. Remember that the PSS representation of the multistationarity region is the yellow region which contains the orange rectangles as well.}
	\label{fig:Gene_Regulatory_BMC_dimer_PSSdegree2_via_Grid}
\end{figure*}

We continue with the example from Section \ref{example:CAD_Finite_Grid}.  Consider the rectangular representation of the multistationarity region of that example given in Figure~\ref{fig:Gene_Regulatory_BMC_dimer_GridRepresentation}. To find the PSS representation of this set, we let $B=[(0.0005,0),(0.0010,2)]$ and $K$ be the union of rectangles $K_i$s such that their associated number is greater than or equal to 2. From the 100 sub-rectangles of $B$, 9 of them satisfy this condition. These sub-rectangles are colored orange in Figure~\ref{fig:Gene_Regulatory_BMC_dimer_PSSdegree2_via_Grid}. We use the \texttt{YALMIP} and \texttt{SeDuMi} packages of Matlab to solve the SOS optimization discussed before this example. To report the computation time we add the two times reported in the output of \texttt{YALMIP}:  the ``\texttt{yalmiptime}" and ``\texttt{solvertime}". It takes about 2 seconds to get the coefficients of the polynomial $p$ of the PSS representation of degree 2. Figure \ref{fig:Gene_Regulatory_BMC_dimer_PSSdegree2_via_Grid} shows the plot of $U(p)$. 

Unfortunately the same code does not produce a better approximation when we increase $d$, the degree of $p$, from 2 to 4, 8 or 16.  The output from Matlab gives similar figures in these cases as Figure~\ref{fig:Gene_Regulatory_BMC_dimer_PSSdegree2_via_Grid}.

\begin{figure*}[p]
	\begin{subfigure}[b]{0.45\textwidth}
		\includegraphics[width=\linewidth]{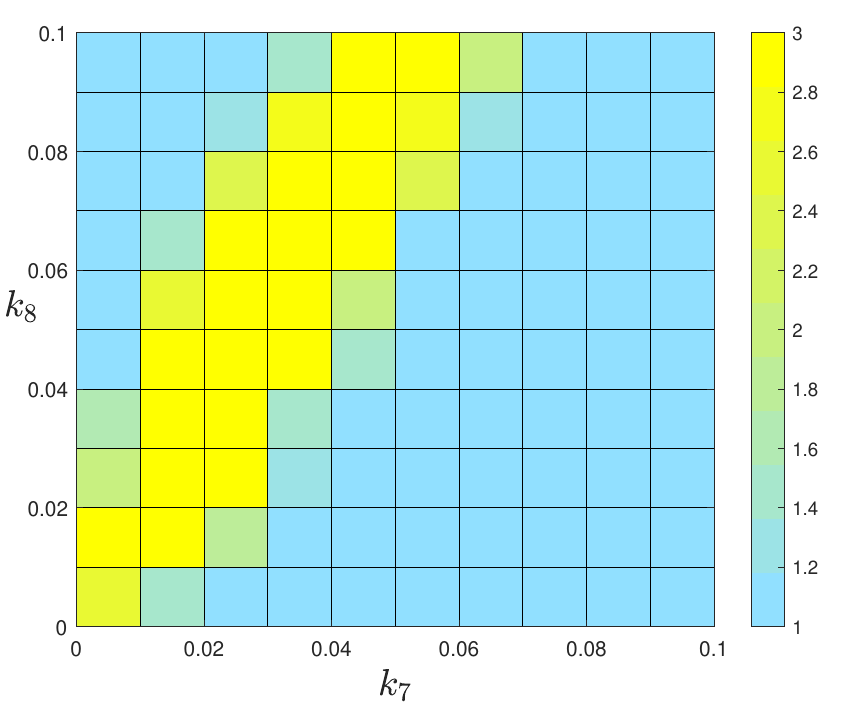}
		\caption{}
		\label{fig:LacI-TetR_mass_action_grid_reproduced}
	\end{subfigure}
	\hfill
	\begin{subfigure}[b]{0.45\textwidth}
		\includegraphics[width=\linewidth]{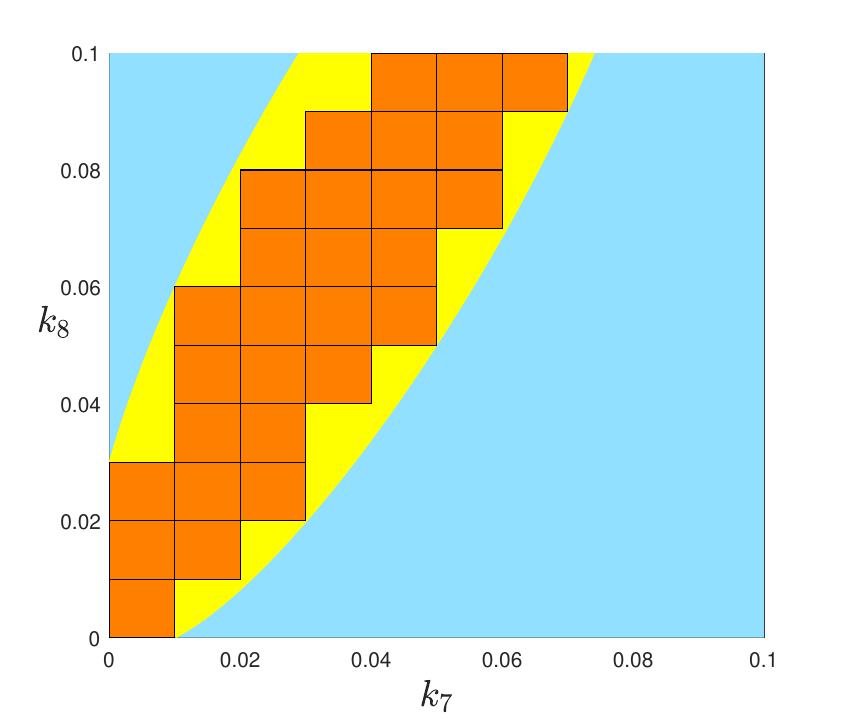}
		\caption{}
		\label{fig:LAcI-TetR_mass_PSS_degree_2_via_grid}
	\end{subfigure}
	\vskip\baselineskip
	\begin{subfigure}[b]{0.45\textwidth}
		\includegraphics[width=\linewidth]{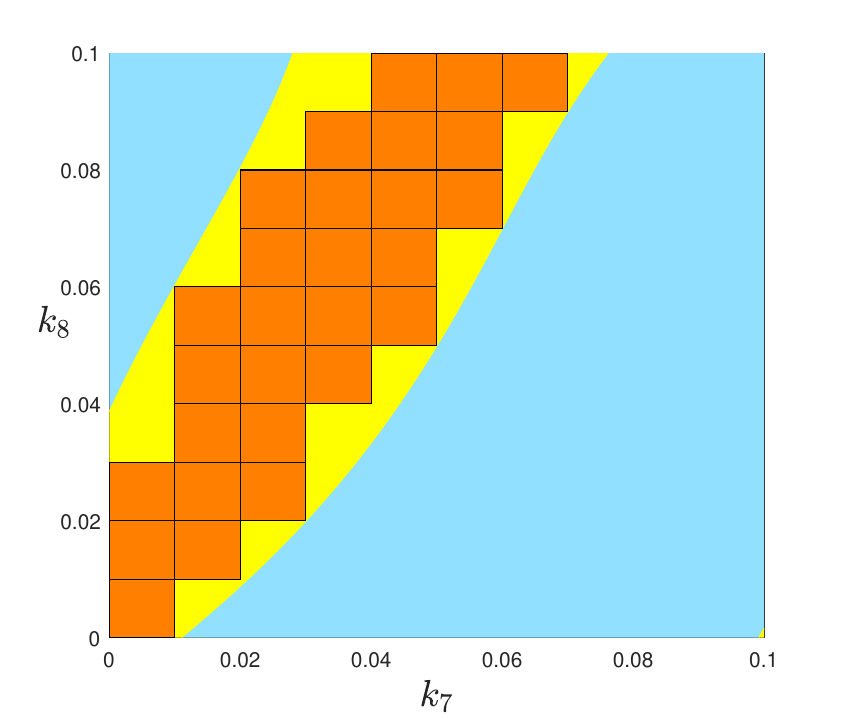}
		\caption{}
		\label{fig:LAcI-TetR_mass_PSS_degree_4_via_grid}
	\end{subfigure}
	\hfill
	\begin{subfigure}[b]{0.45\textwidth}
		\includegraphics[width=\linewidth]{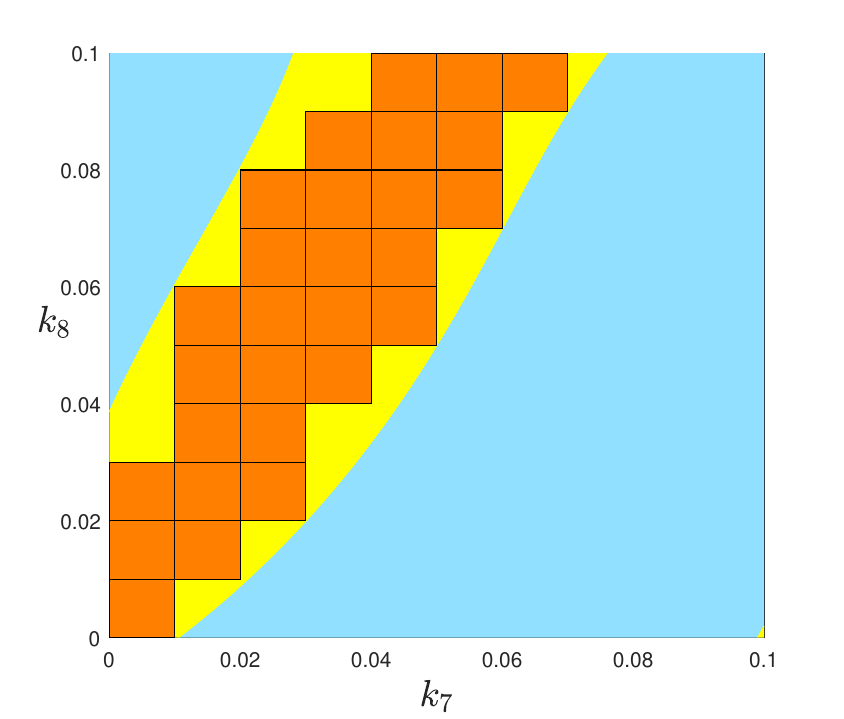}
		\caption{}
		\label{fig:LAcI-TetR_mass_PSS_degree_6_via_grid}
	\end{subfigure}
	\caption{PSS representations of different degrees for the mulistationarity region of the LacI-TetR gene regulatory network using the information we got from the rectangular representation. The union of orange colored subrectangles is considered as the approximation of the multistationarity region obtained by the rectangular representation and chosen as the set $K$. The yellow colored area is the difference of $U(p)-K$. \\
		(a) The rectangular representation of multistationarity region of the network with the system of equations given in \eqref{eq:LacI_TetR_gene_regulatory_network_mass_action_system} and some parameters being fixed by the values in \eqref{eq:Fixed_parameters_of_LacITetR}. 
		\\ (b$-$d) PSS representations of the multistationarity region of degrees 2, 4 and 8 respectively obtained by the information of Figure \ref{fig:LacI-TetR_mass_action_grid_reproduced}.}
	\label{fig:PSS_via_grid}
\end{figure*}

Consider another gene regulatory example from \cite[Chapter 2]{AmirPhDThesis}. To avoid lengthening the text, we only reproduce the system of equations needed to study the multistationarity of the network:
\begin{equation}\label{eq:LacI_TetR_gene_regulatory_network_mass_action_system}
	\begin{array}{ll}
		k_1x_7x_5-k_5x_1=0 &
		\quad k_2x_8x_6-k_6x_2=0\\
		k_3x_1-k_7x_3=0 &
		\quad k_4x_2-k_8x_4=0\\
		k_9x_7x_4-k_{11}x_9=0 &
		\quad k_{10}x_8x_3-k_{12}x_{10}=0\\
		k_{13}x_9x_4-k_{15}x_{11}=0 &
		\quad k_{14}x_{10}x_3-k_{16}x_{12}=0\\
		x_5=k_{17} &
		\quad x_6=k_{18}\\
		x_7+x_9+x_{11}=k_{19} &
		\quad x_8+x_{10}+x_{12}=k_{20}.
	\end{array}
\end{equation}
We fix all parameters other than $k_7$ and $k_8$ to the following values coming from Equation (2.10) of \cite[Chapter 2]{AmirPhDThesis}:
\begin{equation}\label{eq:Fixed_parameters_of_LacITetR}
	\begin{array}{l}
		k_1=k_2=k_3=k_4=1,\;k_5=0.0082,\;k_6=0.0149,\;\\
		k_9=k_{10}=0.01,\;k_{11}=k_{12}=10000,\;k_{13}=2,\;\\
		k_{14}=25,\;k_{15}=1,k_{16}=9,k_{17}=k_{18}=k_{19}=1,\;\\
		k_{20}=4.
	\end{array}
\end{equation}
We reproduced the rectangular representation of the multistationarity region of this network by Matlab, as shown in Figure~\ref{fig:LacI-TetR_mass_action_grid_reproduced}. From the 100 sub-rectangles in total, for 28 of them the average number of steady states is greater than or equal to 2. Using YALMIP and SeDuMi it took between 1 and 2 seconds to get the polynomials of the PSS representations of degrees 2, 4 and 8 represented in Figures \ref{fig:LAcI-TetR_mass_PSS_degree_2_via_grid}, \ref{fig:LAcI-TetR_mass_PSS_degree_4_via_grid} and \ref{fig:LAcI-TetR_mass_PSS_degree_6_via_grid} respectively. For this example, the PSS approximation of degree 4 looks different than that of degree 2, but for degree 8 the plot  looks similar to degree 4.

\subsection{Advantages of PSS representation over rectangular representation}
\label{remark:Comparision-PSS-grid}

It is natural to ask why one should find a PSS representation of the multistationarity region using the rectangular representation given one already has the rectangular representation? Let $B\subseteq\R^r$ be the parameter region of the form of a hyperrectangle, and $K\subseteq B$ be the multistationarity region. In the rectangular representation we have $K\simeq\cup_{i=1}^m K_i$ where $K_i=[a_{K_i},b_{K_i}]$ are hyperrectangles. In the PSS representation we have $K\simeq U(p)$ where $p$ is a polynomial of degree $d$.
\begin{itemize}
	\item[1-] When $r\geq 4$, plotting $K$ is impossible. In order to save or show the rectangular representation one needs to use a matrix of size $(m)\times (2r)$, where each row stands for one $K_i$ and the first $r$ columns have the coordinates of the point $a_{K_i}$ and the second $r$ columns correspond to the coordinates of the point $b_{K_i}$. However for the PSS representation one needs to use only a vector of size $\binom{r+d}{r}=\sum_{i=0}^d\binom{r-1+i}{r-1}$, where $\binom{r-1+i}{r-1}$ entries are coefficients of the terms of degree $i$. The terms are ordered from smaller total degree to larger and for the terms of the same total degree we use the lexicographic order.
	\item[2-] To test if a point $k^\star\in B$ belongs to $K$ using the rectangular representation one should check $m$ conditions of the form $k^\star\in K_i$ which means verifying an inequality on each coordinate of the point, i.e. $a_{K_i,j}\leq k^\star_j\leq b_{K_i,j}$. If one of the conditions $k^\star\in K_i$ is positive, then there is no need to check the rest, otherwise all should fail to conclude that $k^\star\not\in K$. However, using the PSS representation one needs to check only one condition of an evaluation form, i.e. $p(k^\star)\geq 1$.
	\item[3-] Recall from the last paragraph of Section \ref{ssec:sampling} explaining that parameters near the boundary of the multistationarity region are not suitable choices for an experimentalist. To check the distance of a point $k^\star\in B\setminus K$ to the boundaries of $K$ using the rectangular representation one should find distance of $k^\star$ from boundaries of each $K_i$ and then taking the minimum. However, using the PSS representation, in both cases of $k^\star\in B\setminus K$ or $k^\star\in K$, one just needs to find the distance of $k^\star$ from the algebraic set defined by $p(k)-1=0$, for example by Lagrange multipliers, as in the next section.
\end{itemize}

To conclude, if $\binom{r+d}{r}$ is considerably smaller than $2mr$, then storing the PSS representation instead of the initial rectangular representation will save memory without loosing information about the multistationarity region.

\FloatBarrier

\subsection{Approximating the distance of parameter point from the boundary}
\label{example:Lagrange_multipliers}

To illustrate how to approximate distance of a parameter point from the boundaries of the multistationarity region using a PSS representation we continue with the example from Sections \ref{example:Comparision-PSS-grid} and \ref{example:CAD_Finite_Grid}.

Let $p$ be the polynomial of degree 2 in two variables $k_7$ and $k_8$ corresponding to $U(p)$ in Figure~\ref{fig:LAcI-TetR_mass_PSS_degree_2_via_grid}. We will approximate distance of the point $k^\star=(0.08,0.02)$ from the boundary of the multistationarity region by the distance of $k^\star$ from the algebraic set defined by $p(k_7,k_8)-1=0$. This question is equivalent to minimizing the Euclidean distance function of a point $(k_7,k_8)$ from the point $k^\star$ subject to the constraint $(k_7,k_8)\in L_1(p)$. The target function is $\sqrt{(k_7-0.08)^2+(k_8-0.02)^2}$ which gets minimized if and only if $(k_7-0.08)^2+(k_8-0.02)^2$ gets minimized. An elementary way to solve this minimization problem is to use the method of Lagrangian multipliers \cite[Chapter 7, Theorem 1.13]{Fuente_2000}. Define
\begin{equation*}
	F(k_7,k_8,\lambda) = (k_7-0.08)^2+(k_8-0.02)^2
	+ \lambda\big(p(k_7,k_8)-1\big).
\end{equation*}
Now we must find the critical points of $F(k_7,k_8,\lambda)$. So we should solve the system of equations obtained by $\frac{\partial F}{\partial k_7}=\frac{\partial F}{\partial k_8}=\frac{\partial F}{\partial \lambda}=0$. It takes 0.167 seconds to solve this system of equations by the \texttt{solve} command in Maple. It has 4 solutions, from which 2 of them belong to the rectangle $B=[(0,0),(0.1,0.1)]$ and the minimum of the target function is obtained at the point $(0.04499222669,0.04161251428)$. The distance of this point from $k^\star$ is $0.04114176669$.

\subsection{Constructing a PSS representation from a sampling representation}
\label{example:Comparision-PSS-finite}

It is not necessary to have a rectangular representation to get the PSS representation. Let $B=[a_B,b_B]$ be a hyperrectangle and $K=\{a^{(1)},\dots,a^{(m)}\}$ a finite set. Let $d\in\N$, and the goal be to find the coefficients of a polynomial of degree $d$ such that $\int_Bp(x)dx$ becomes minimum subject to some conditions. We already saw that the target function is linear. The constraint $p\geq 1$ on $K$ can be enforced by $p(a^{(i)})\geq 1$ for every $i$, which are linear constraints. The positivity of $p$ on $B$ can be enforced by Equation \eqref{eq:constraints_nonnegative_on_hyperrectangle} or by adding a large enough number of random points from $B$ and putting the constraint $p(a)>0$. The later idea makes the problem solvable by any common linear programming tool. However, here we still use Equation \eqref{eq:constraints_nonnegative_on_hyperrectangle}. 

\begin{figure*}[ht]
	\centering
	\begin{subfigure}[b]{0.29\textwidth}
		\includegraphics[width=\linewidth]{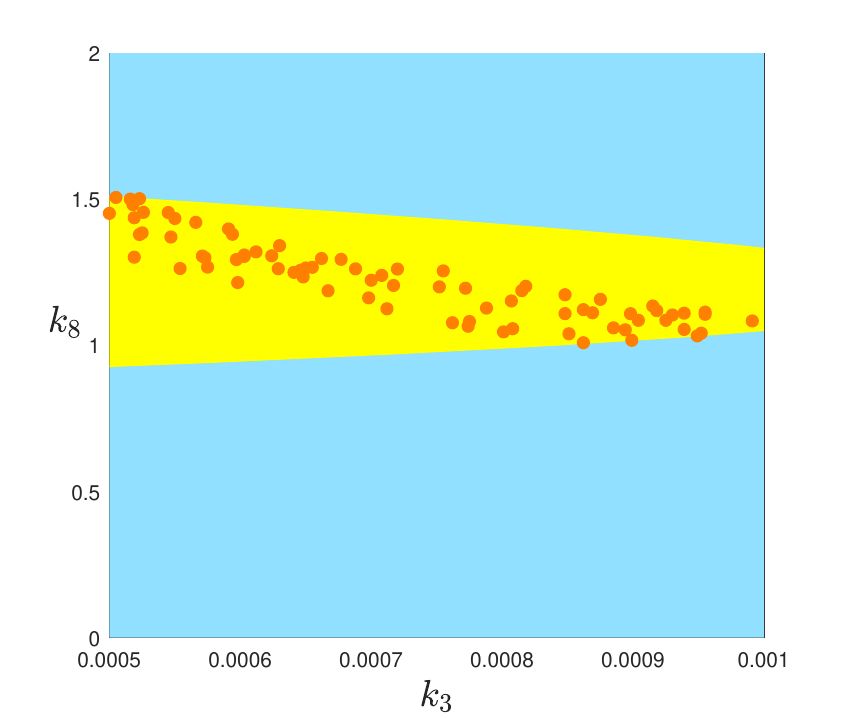}
		\caption{}
		\label{fig:Gene_Regulatory_BMC_dimer_PSS_d_2_via_finite}
	\end{subfigure}
	\hfill
	\begin{subfigure}[b]{0.29\textwidth}
		\includegraphics[width=\linewidth]{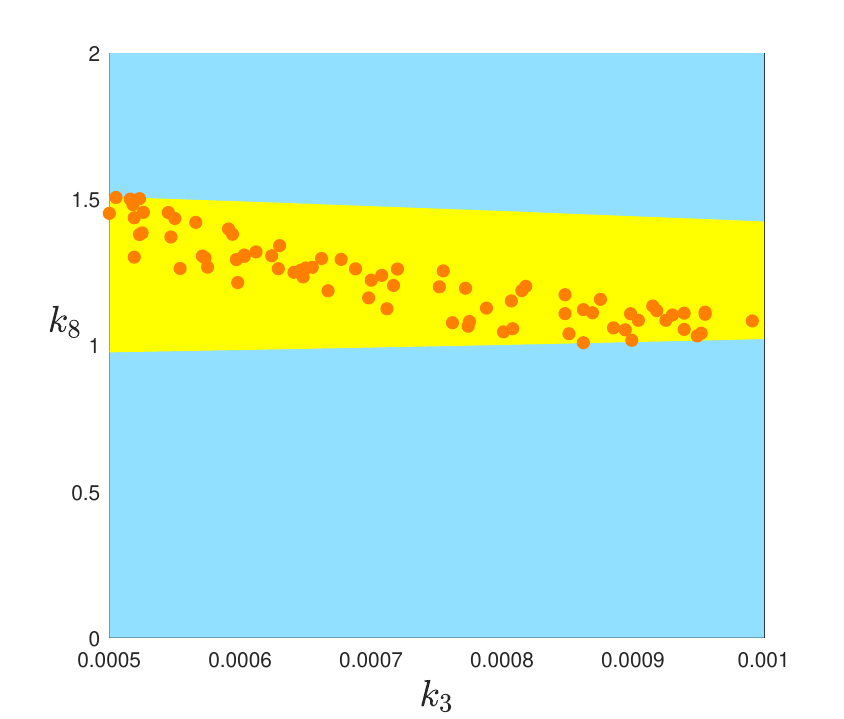}
		\caption{}
		\label{fig:Gene_Regulatory_BMC_dimer_PSS_d_6_via_finite}
	\end{subfigure}
	\hfill
	\begin{subfigure}[b]{0.29\textwidth}
		\includegraphics[width=\linewidth]{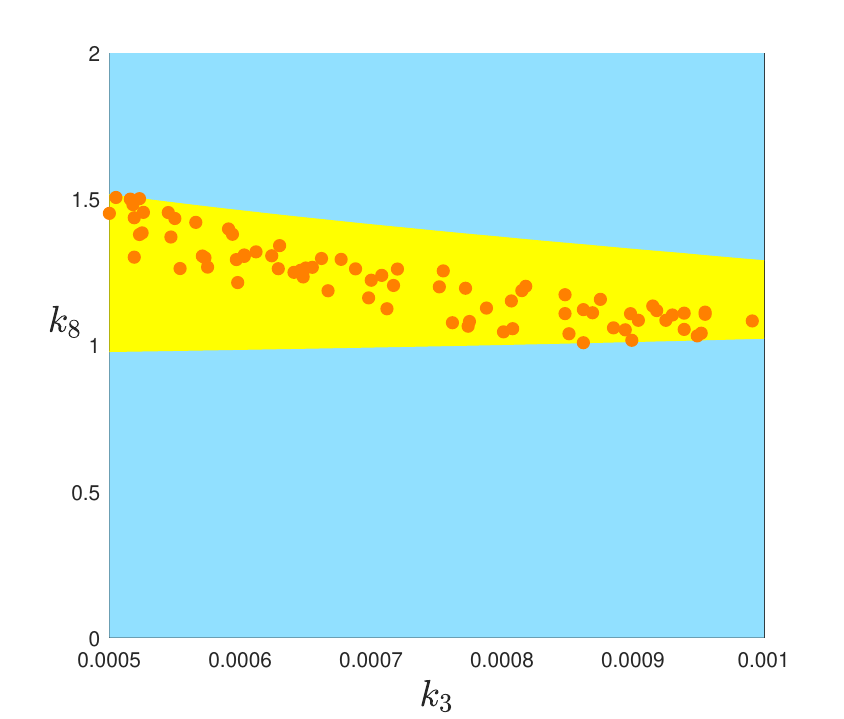}
		\caption{}
		\label{fig:Gene_Regulatory_BMC_dimer_PSS_d_10_via_finite}
	\end{subfigure}\\
	\begin{subfigure}[b]{0.29\textwidth}
		\includegraphics[width=\linewidth]{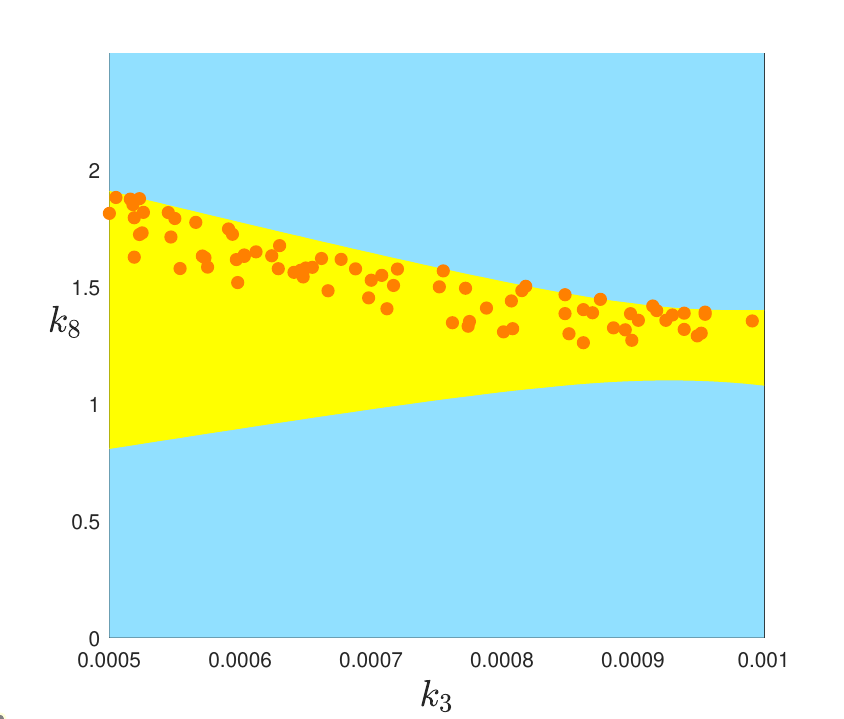}
		\caption{}
		\label{fig:Gene_Regulatory_BMC_dimer_PSS_d_2_via_finite_re}
	\end{subfigure}
	\hfill
	\begin{subfigure}[b]{0.29\textwidth}
		\includegraphics[width=\linewidth]{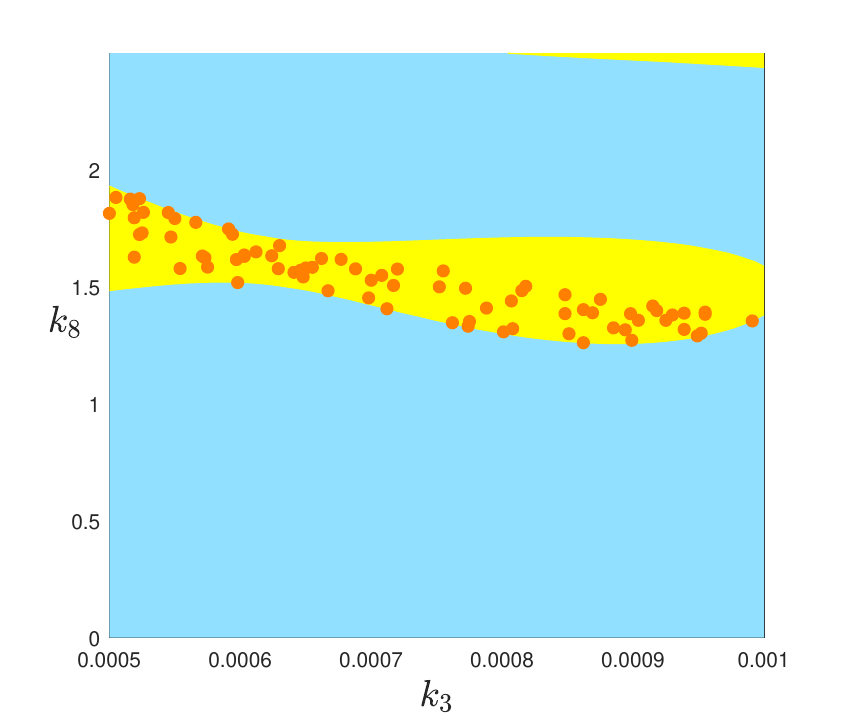}
		\caption{}
		\label{fig:Gene_Regulatory_BMC_dimer_PSS_d_6_via_finite_re}
	\end{subfigure}
	\hfill
	\begin{subfigure}[b]{0.29\textwidth}
		\includegraphics[width=\linewidth]{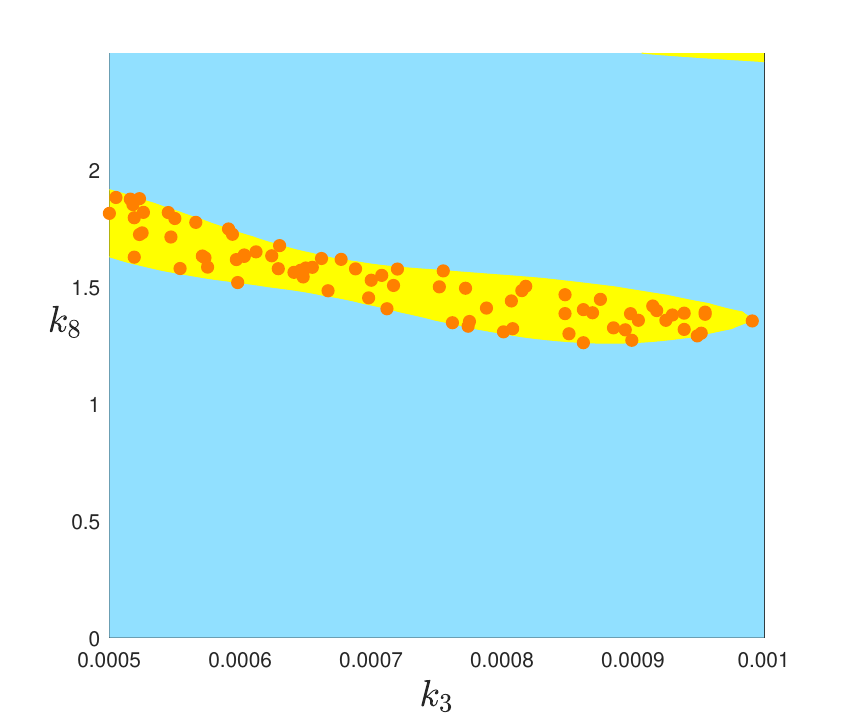}
		\caption{}
		\label{fig:Gene_Regulatory_BMC_dimer_PSS_d_10_via_finite_re}
	\end{subfigure}
	\caption{PSS representation of different degrees of the mulistationarity region of the network of Example~\ref{example:CAD_Finite_Grid} inside the hyperrectangle $B=[(0.0005,0),(0.001,2)]$ using the information we got from the sampling representation of the multistationairy region. The orange colored points are the points with three steady states and their union is considered as approximation of $K$. The yellow colored area is the difference of $U(p)-K$. One expects to see that this difference is getting smaller as the degree increases. However, the Matlab code that we wrote using YALMIP and SeDuMi does not behave as expected. \\
		(a)$-$(c) gives the PSS representation of the original problem of degrees 2, 6 and 10 respectively.  \\
		(d)$-$(f) gives the PSS representation of those degrees for the problem after after rescaling the parameters for better numerical behavior via \texttt{YALMIP} and \texttt{SeDuMi}.}
	\label{fig:LacI-TetR-PSSRepresentation-2}
\end{figure*}

Let us illustrate this with our ongoing example.  Consider the sampling representation of the multistationarity region of the network of Example~\ref{example:CAD_Finite_Grid} given at Figure~\ref{fig:Gene_Regulatory_BMC_dimer_FiniteRepresentation}. To find the PSS representation of this set, we let $B=[(0.0005,0),(0.001,2)]$ and $K$ to be the set of points for which the system $f_k(x)=0$ had more than one positive solution. There are 1000 points from which 78 of them are parameter choices where the network has three steady states. Using the \texttt{YALMIP} package of Matlab, it takes less than a second to get the coefficients of each of the polynomials $p$ of the PSS representation of degrees 2, 6 and 10. Figures~\ref{fig:Gene_Regulatory_BMC_dimer_PSS_d_2_via_finite}, \ref{fig:Gene_Regulatory_BMC_dimer_PSS_d_6_via_finite} and \ref{fig:Gene_Regulatory_BMC_dimer_PSS_d_10_via_finite} show the plots of $U(p)$ for degrees 2, 6 and 10 respectively. The plot for degree 6 actually looks worse than the plot for degree 2 (further away from the actual result in Figure \ref{fig:Gene_Regulatory_BMC_dimer-CAD}), although the one for degree 10 looks a little better.  In all these cases \texttt{YALMIP} finished the computations with a message \texttt{`Numerical problems (SeDuMi)'} indicating that the solver found the problem to be numerically ill-posed. Rescaling the parameter region of interest, $B$, to $[(0,0),(1,1)]$ and then transforming the PSS polynomial back to the original $B$ allows a better PSS approximations via \texttt{YALMIP}. For degrees 2 and 6 the numerical problem message is avoided but for degree 10 it remains. The results are shown in Figures~\ref{fig:Gene_Regulatory_BMC_dimer_PSS_d_2_via_finite_re}, \ref{fig:Gene_Regulatory_BMC_dimer_PSS_d_6_via_finite_re} and \ref{fig:Gene_Regulatory_BMC_dimer_PSS_d_10_via_finite_re}.

To demonstrate that the number of free parameters need not be 2 to be able to compute the PSS representation, we repeated the above process with all 8 parameters of the system being free in the following hyperrectangle:
\begin{equation*}
	B = [\, (1, 0, 0.0005, 0, 1, 1, 40, 0),
	(4, 2, 0.001, 2, 4, 3, 50, 2) \,].
\end{equation*}
Solving the system at 1000 random points uniformly chosen from $B$ takes the same amount of time as solving the system at 1000 random points with only 2 of their coordinates varying took time in the previous case. It took about 1.5 seconds to compute the 45 coefficients of the polynomial of the PSS representation of degree 2. The polynomial found is the following.
\begin{align*}
	p &= 
	0.0232593410 k_{1}^{2}- 0.2965863774 k_{2}^{2}+ 559.9760177000 k_{3}^{2}- 0.0618678914 k_{4}^{2}+\\
	& 0.0098179292 k_{5}^{2}- 0.0061375489 k_{6}^{2}- 0.0036871825 k_{7}^{2}- 0.0716829116 k_{8}^{2}-\\
	& 0.0110701085 k_{1} k_{2}- 5.3373104650 k_{1} k_{3}+ 0.0186292240 k_{1} k_{4}+ 12.5539028600 k_{2} k_{3}+\\
	& 0.0051935911 k_{1} k_{5}- 0.0820767495 k_{2} k_{4}+ 0.0250525965 k_{1} k_{6}- 0.0649618436 k_{2} k_{5}-\\
	& 8.2165668850 k_{3} k_{4}+ 0.0092355627 k_{1} k_{7}- 0.0310047257 k_{2} k_{6}- 4.2926137530 k_{3} k_{5}-\\
	& 0.0527315151 k_{1} k_{8}+ 0.0017547696 k_{2} k_{7}- 2.7647028150 k_{3} k_{6}+ 0.0288291263 k_{4} k_{5}+\\
	& 0.0785940967 k_{2} k_{8}- 0.4049844472 k_{3} k_{7}+ 0.0536360086 k_{4} k_{6}+ 12.6014929300 k_{3} k_{8}-\\
	& 0.0091147734 k_{4} k_{7}+ 0.0198800552 k_{5} k_{6}- 0.0567174819 k_{4} k_{8}+ 0.0009448114 k_{5} k_{7}-\\
	& 0.0190068615 k_{5} k_{8}+ 0.0102920949 k_{6} k_{7}+ 0.0316830973 k_{6} k_{8}+ 0.0022468043 k_{7} k_{8}-\\
	& 0.6043381476 k_{1}+ 0.7152344469 k_{2}+ 0.3287484744 k_{4}+ 36.3811699100 k_{3}-\\
	& 0.0617318336 k_{5}- 0.5940226892 k_{6}+ 0.2987521874 k_{7}+ 0.2558883329 k_{8}- 5.0441247030.
\end{align*}

\subsection{Advantages of PSS representation over the sampling representation}
\label{remark:Comparision-PSS-finite}

A similar question to that in Section \ref{remark:Comparision-PSS-grid} can be asked here: why should one find a PSS representation of the multistationarity region using the sampling representation if he already has a sampling representation? Let $B\subseteq\R^r$ be the parameter region of the form of a hyperrectangle, $K\subseteq B$ be the multistationarity region. In the sampling representation we have $\{a^{(1)},\dots,a^{(m)}\}\subseteq K$. In the PSS representation we have $K\simeq U(p)$ where $p$ is a polynomial of degree $d$.
\begin{itemize}
	\item[1-] When $r\geq 4$, plotting $K$ is impossible. In order to save or show the sampling representation one needs to use a matrix of the size $m \times r$, where each row stands for one point $a^{(i)}$ and the columns correspond to the coordinates of the points. However, for the PSS representation one needs to use a vector of the size $\binom{r+d}{r}$, as explained in Section~\ref{remark:Comparision-PSS-grid} item 1.
	\item[2-] To test if a point $k^\star\in B$ belongs to $K$ using the sampling representation is not a straightforward task. However, using the PSS representation one needs to verify only one condition of the evaluation form, $p(k^\star)\geq 1$.
	\item[3-] To compute the distance of a point $k^\star\in B$ to the boundaries of $K$, using the sampling representation, if $k^\star\not\in K$, one should compute the distance of $k^\star$ from each point in the sampling representation of $K$ and then take the minimum. Using the PSS representation, whether $k\not\in K$ or not, one just needs to find the distance of $k^\star$ from the algebraic set defined by $p(k)-1=0$.
\end{itemize}
In a typical example from CRN theory, $r$ is usually much higher than 2, and therefore item 1 is really important. When $\binom{r+d}{r}$ is lower than $rm$, one can use less memory by saving the PSS representation instead of keeping all the points of the sampling representation in the memory. Further, as items 2 and 3 show, this will not cause a loss of information about the multistationarity region.

\subsection{More involved example}
\label{example:S-system}

We already showed at the end of Section~\ref{example:Comparision-PSS-finite} that the PSS representation can be generated for examples with a higher number of parameters than two. That is, we let all 8 parameters of the network in Figure~\ref{fig:Gene_Regulatory_BMC_dimer-network} to be free and found the PSS representation of degree two in 8 indeterminants. Now we bring annothr such example which also serves to emphasize Remark~\ref{remark:conservation-laws-and-kinetics-choice} item (ii): that to have a PSS representation of the multistationarity region, one does not need to have the right hand side functions of the ODE system to be of polynomial or even rational functions.

\begin{figure*}[ht]
	\centering
	\begin{subfigure}[b]{0.52\textwidth}
		\vspace{-1cm}
		\centering
		
		\resizebox{\linewidth}{!}{
			\begin{minipage}{\linewidth}
				\begin{align*}
					\frac{dx_1}{dt} &= \alpha_1\Big(v_1+\frac{v_2x_4^n}{v_3^n+x_4^n}\Big)\cdot x_3^{-0.5} -\beta_1x_1^{h_{1,1}},\\ 
					\frac{dx_2}{dt} &= \alpha_2x_1^{g_{2,1}} -\beta_2x_2^{h_{2,2}},\\
					\frac{dx_3}{dt} &= \alpha_3x_2^{g_{2,2}} -\beta_3x_3^{h_{3,3}},\\ \frac{dx_4}{dt} &= \alpha_4x_3^{g_{4,3}} -\beta_4x_4^{h_{4,4}}.	
				\end{align*}
				
			\end{minipage}
		}
		\caption{}
		\label{fig:S_system_ODE}
	\end{subfigure}
	\hfill
	\begin{subfigure}[b]{0.4\textwidth}
		\centering
		\includegraphics[width=5cm]{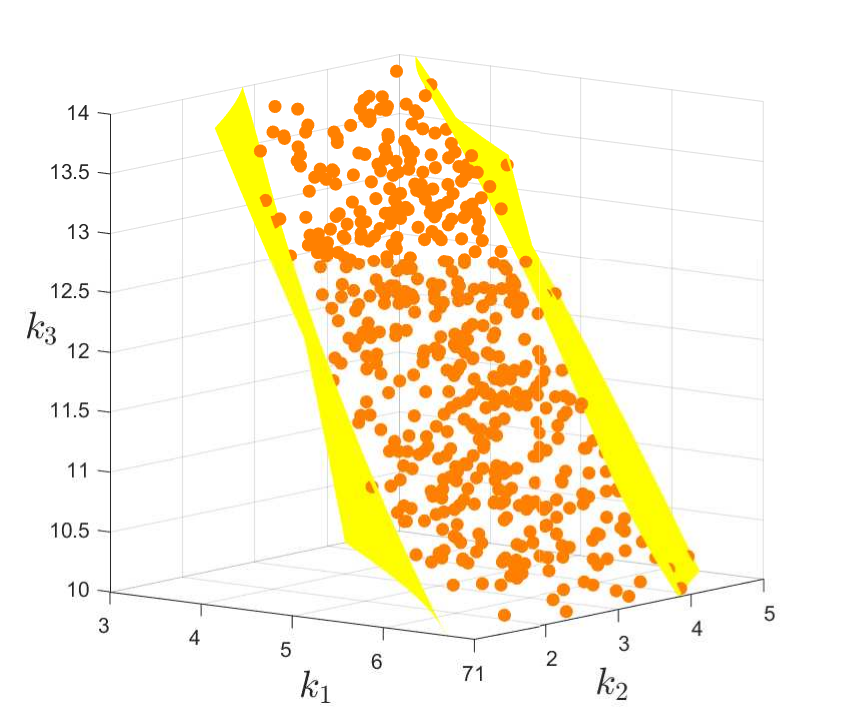}
		\caption{}
		\label{fig:S_system_PSS_representation_d_2}
	\end{subfigure}
	\caption{A gene expression network with an S-system kinetics that allows Hill terms and exhibits multistationary behavior, taken from \cite[Figure 2]{Voit-2005}. (a)The ODE system of the network. The set of species is $\lbrace X_1, X_2, X_3, X_4\rbrace$. Each equation $dx_i/dt$ consists of two terms responsible for production and degradation of the species $X_i$. The right hand side functions are not necessarily of a polynomial type, or even a rational function form. (b) The sampling and PSS representation of the multistationarity region of this network when all the parameters other than $\beta_i$, $i=2,3,4$ are fixed to the values in \eqref{eq:Fixed_parameters_of_S-system} and the remaining three parameters renamed $k_i$, $i=1,2,3$ vary in the cube $B=[(3,1,10),(7,5,14)]$. The sampling representation consists of the orange color points and the PSS representation is the area between the two yellow colored surfaces.}
	\label{fig:S_system}
\end{figure*}

Consider a gene expression system with 4 species $X_i$, $i=1,\dots,4$ where these species can be m-RNA or protein molecules or other relevant factors, with the ODE system as in Figure~\ref{fig:S_system_ODE} which was introduced in \cite[Figure 2]{Voit-2005}. As one can see the right hand side functions involve at least a square root, and as a result this system is not polynomial, or even defined by rational functions. Let us fix the parameter values other than the three degradation rates $\beta_i$, $i=2,3,4$ to the following values, chosen the same as in \cite{Voit-2005}:
\begin{equation}\label{eq:Fixed_parameters_of_S-system}
	\begin{array}{l}
		\alpha_1=1,\;n=4,\;v_1=4,\;v_2=8,\;v_3=4,\;\\
		h_{1,1}=h_{2,2}=0.5,\;h_{3,3}=1,\;h_{4,4}=0.75,\;\\
		g_{2,1}=g_{3,2}=g_{4,3}=1,\;\beta_1=0.5,\;\\
		\alpha_2=1,\;\alpha_3=2,\;\alpha_4=3.
	\end{array}
\end{equation}

In \cite{Voit-2005} it is shown that for $(\beta_2,\beta_3,\beta_4)=(5,3,12)$, the network is bistable, with three steady states. Here we rename these three parameters by $k_i$, $i=1,2,3$ respectively, let them vary in the 3-dimensional cube $B=[(3,1,10),(7,5,14)]$, and look for the multistationary region. I.e. we seek the relation between these three parameters so that the number of steady states of the network remains the same. Substituting the values into Figure~\ref{fig:S_system_ODE} and letting $dx_i/dt=0$, $i=1,\dots,4$ gives the following system of equations.
\begin{equation}\label{eq:S-system_equations}
	\begin{array}{l}
		4+\frac{8x_4^4}{256+x_4^4}\frac{1}{\sqrt{x_1}}-\frac{1}{2}\sqrt{x_1}=0,\\
		x_1-k_1\sqrt{x_2}=0,\\
		2x_2-k_2x_3=0,\\
		3x_3-k_3\sqrt[4]{x_4^3}=0.
	\end{array}
\end{equation}
With a simple calculation one can see that the set of positive solutions to the system of equations \eqref{eq:S-system_equations} is in one to one correspondence with the the set of positive roots of the following univariate polynomial:
\begin{equation}\label{eq:S-system_main_polynomial}
	f(y) = \sqrt[4]{24k_1^2k_2k_3^3}y^{73} - 144y^{64} + 256\sqrt[4]{24k_1^2k_2k_3^3}y^9 - 12288.
\end{equation}
For any positive root of the polynomial in \eqref{eq:S-system_main_polynomial}, a positive solution for \eqref{eq:S-system_equations} is 
\begin{equation*}
	(x_1,x_2,x_3,x_4) =
	\Big(k_1\sqrt{\frac{k_2k_3}{6}}y^6,\,\frac{k_2k_3}{6}y^{12},\frac{k_3}{3}y^{12},y^{16}\Big).
\end{equation*}
By Descartes rule of signs it is clear that the polynomial in \eqref{eq:S-system_main_polynomial} has 1 or 3 positive roots counted by multiplicity for any choice of $k_i$s. To find the sampling representation of the multistationarity region of our system we solved the equation $f(y)=0$ for 1000 random choices of the parameters using Matlab, which took about 8 minutes. At 532 of these choices the polynomial had 3 positive roots. The sampling representation is shown in Figure~\ref{fig:S_system_PSS_representation_d_2}, with the points where the system has 3 steady states shown as orange spheres. We then used the information from the sampling representation to compute the PSS representation of the multistationarity region of degree two, which took less than 2 seconds. The resulting polynomial is below, with coefficients given to 6 decimal places.
\[
\begin{array}{l}
	p(k_1,k_2,k_3) = - 0.043653k_1^2 - 0.029919k_2^2 \\
	-0.013650k_3^2 - 0.082681k_1k_2 - 0.045005k_1k_3 \\ -0.055895k_2k_3 + 1.226648k_1 + 1.271835k_2 \\ 
	+0.710556k_3 - 8.112402.
\end{array}
\]
Therefore the PSS representation of the multistationarity region is the set of the points $(k_1,k_2,k_3)\in B$ that satisfy $p(k_1,k_2,k_3)\geq 1$, which is the region between the two yellow surfaces in Figure~\ref{fig:S_system_PSS_representation_d_2}. Note that the point $(5,3,12)$ corresponding to the value examined in \cite{Voit-2005} is in the middle of this region.

\section{Bisection algorithms for rectangular representation}
\label{sec:Bisect search}

As shown in the previous section, it is possible to get the PSS representation both from the sampling and the rectangular representations.
If one needs to solve the system at random points and take an average in order to get the rectangular representation, then using the rectangular representation does not have much advantage over using the sampling representation for the purpose of finding the PSS representation.

But in some cases it is possible to compute the average number of the solutions without solving the system. One such case is introduced in \cite{Feliu-Sadeghimanesh-2020}. Instead of solving the system for many points, it is enough to compute one integral called the Kac-Rice integral. In this situation, if the computation of the integral is possible and faster than solving the system for many random points, then the rectangular representation can be preferred to the sampling representation. However, one still needs a computation per each sub-hyperrectangle of the rectangular representation and this number can grow by the number of parameters. If $B\subseteq\R^r$ and we divide it along each axis to $m$ equal parts, then the number of sub-hyperrectangles in the rectangular representation becomes $m^r$. A different approach to build a rectangular decomposition is to use bisection algorithms instead of dividing $B$ to equal sub-hyperrectangles. When using a bisection algorithm, the rectangular decomposition usually contains fewer number of sub-hyperrectangles (not necessarily of equal volume).

\subsection{Two possible number of solutions for given parameter point}

Let us simplify the question. There is a hyperrectangle $B\subseteq\R^r$ and a function $g\colon B\rightarrow\Z_{\geq 0}\cup\{\infty\}$ which in our case is $\Phi_f^0$ associated to a parametric function $f_k(x)$. Let $\mathcal{B}$ be the set containing all sub-hyperrectangles of $B$. The goal is to express $L_i(g)$ or $U_2(g)$ as a union of sub-hyperrectangles of $B$. 
One of the shapes of multistationary networks, observed often for realistic models, are bistable networks with a folding type of bifurcation, such as in dual phosphorylation cycle \cite{Markevich-Hoek-Kholodenko-2004}.

In our settings these networks have one steady state for some choices of parameters and three\footnote{Two stable and one unstable, however we do not mention stability of the steady states in this paper.} steady states for some other choices of the parameters and for a zero measure set of parameters in the boundary of the two regions it has two steady states. The networks in Sections \ref{example:CAD_Finite_Grid} and \ref{example:Comparision-PSS-grid} are examples of such networks. In such cases $\Phi_f^0$ is almost always either 1 or 3. Going back to our question, motivated from application, assume $\Image(g)=\{n_1,n_2\}$ where $n_1\lneqq n_2$. In this case for each $S\in\mathcal{B}$ one of the followings occurs.
\begin{itemize}
	\item[i)] $\mathbb{E}\big(g(k)\mid k\sim U(S)\big)=n_1.$\\
	This can happen if and only if for almost every $k\in S$, $g(k)=n_1$. 
	\item[ii)] $\mathbb{E}\big(g(k)\mid k\sim U(S)\big)=n_2$\\
	This can happen if and only if for almost every $k\in S$, $g(k)=n_2$.
	\item[iii)] $\mathbb{E}\big(g(k)\mid k\sim U(S)\big)=\alpha$, $n_1\lneqq \alpha\lneqq n_2$.\\
	This can happen if and only if $S\cap L_{n_1}(g)$ and $S\cap L_{n_2}(g)$ both are nonempty and of non-zero measure.
\end{itemize}
The proof is straightforward by noting that
\begin{equation*}
	\mathbb{E}\big(g(k)\mid k\sim U(K)\big)=n_1\overline{\Vol}\big(K\cap L_{n_1}(g)\big)+
	n_2\overline{\Vol}\big(K\cap L_{n_2}(g)\big).
\end{equation*}
Therefore one can compute $\mathbb{E}\big(g(k)\mid k\sim U(B)\big)$. Then if the answer is $n_1$ conclude that almost the whole $B$ is a subset of $L_{n_1}(g)$ and if the answer is $n_2$, conclude that almost the whole $B$ is a subset of $L_{n_2}(g)$. Otherwise we proceed by dividing $B$ along only one axis into two equal sub-hyperrectangles. We continue in this fashion until each sub-hyperrectangle is inside $L_{n_1}(g)$ or $L_{n_2}(g)$ or a termination condition on the length of the edges of the sub-hyperrectangles is met. When the termination condition on the edges is obtained, we put the sub-hyperrectangle in $L_{n_i}(g)$ if $\mathbb{E}(g(k))$ on this sub-hyperrectangle is closer to $n_i$. We refer to this approach as the \emph{two-value bisection search} hereafter. This algorithm is formally written in Algorithm~\ref{Algorithm:bisect-search-two-value}. 

\IncMargin{1em}
\begin{algorithm}[ht]
	\SetKwInOut{Input}{Input}
	\SetKwInOut{Output}{Output}
	\Input{$B=[a,b]\subseteq\R^r$, $g\colon B\rightarrow\{n_1,n_2\}\subseteq\Z_{\geq 0}$, $n_1\lneqq n_2$, $\epsilon\in\R_{>0}$.}
	\Output{$L_{n_1}(g)\simeq\cup_{i=1}^{m_1}S_{n_1,i}$, $L_{n_2}(g)\simeq\cup_{i=1}^{m_2}S_{n_2,i}$, $\forall i,j\colon S_{n_i,j}\in\mathcal{B}$ and the minimum of the length of the edges of $S_{n_i,j}>\epsilon$.}
	\BlankLine
	\emph{initialization}:
	$L_1=\{\}$, $L_2=\{\}$, $T=\{(B,1)\}$\;
	\While{$T\neq\emptyset$}{choose the first element of $T$ and call its first element by $S$ and its second element by $i$ and remove $(S,i)$ from $T$\;
		compute $\alpha=\mathbb{E}\big(g(k)\mid k\sim U(S)\big)$\;
		\Switch{the value of $\alpha$}{
			\Case{$n_1$}{add $S$ to $L_1$\;}
			\Case{$n_2$}{add $S$ to $L_2$\;}
			\Other{define $\beta$ to be the minimum of the length of the edges of $S$\;
				\eIf{$\beta\leq\epsilon$}{
					\eIf{$\alpha\leq\frac{n_1+n_2}{2}$}{add $S$ to $L_1$\;}{add $S$ to $L_2$.}}{define $S_1$ and $S_2$ by dividing $S$ along the $i$-th axis to two equal sub-hyperrectangles. Replace $i$ by $(i+1)\mod n$ and add $(S_1,i)$ and $(S_2,i)$ to $T$\;}
			}
		}
	}
	\caption{A bisection algorithm for building a rectangular representation of a parameter region when there are only two possible number of solutions for the system of equations,  ignoring a zero measure set of parameter values.}\label{Algorithm:bisect-search-two-value}
\end{algorithm}\DecMargin{1em}

\FloatBarrier

A similar algorithm was presented in \cite[Section 2]{Feliu-Sadeghimanesh-2020}. The first difference is that the input to the bisection algorithm in \cite{Feliu-Sadeghimanesh-2020} is not necessarily a system with two general number of solutions. The second difference is that the output, there, is not only the two lists $L_{n_1}(g)$ and $L_{n_2}(g)$ (compare Figure~\ref{fig:HK}(d-f) with \cite[Figure~1c]{Feliu-Sadeghimanesh-2020}).\medskip

If the length of the edges of $B$ are of different scales, then it is better to replace the termination condition of Algorithm~\ref{Algorithm:bisect-search-two-value} with the following:
\[
\min\Big\lbrace\frac{b_{S,j}-a_{S,j}}{b_{B,j}-a_{B,j}}\mid 1\leq j\leq r\Big\rbrace\leq\epsilon,
\]
where $S=[a_S,b_S]$, $B=[a_B,b_B]$ and $0<\epsilon<1$.  The motivation is that one usually is interested in knowing the parameter values with some number of digits of accuracy, after writing the number in scientific notation.

We are not going to explain how to use the Kac-Rice integral in the CRN framework as it is not the topic of this paper. All algorithms in this paper are independent from the choice of the algorithm to compute the expected number of solutions of a parametric system on a given parameter region. A simple algorithm to achieve this is to solve the system for several random points from the given hyperrectangle according to a random distribution of interest and take the average. Thus we may assume the existence of a method capable of computing $\mathbb{E}\big(\Phi_f^0(k)\mid k\sim q\big)$ where $q$ is a distribution on $S$ and then find a rectangular representation using two-valued bisection search and afterwards a PSS representation.

\subsection{Example}

\begin{figure*}[p]
	\centering
	\begin{subfigure}[b]{0.32\textwidth}
		\centering
		\resizebox{\linewidth}{!}{
			\begin{minipage}{\linewidth}
				\[\begin{array}{c}
					X_1\ce{->[k_1]}X_2\ce{->[k_2]}X_3\ce{->[k_3]}X_4\\
					X_3+X_5\ce{->[k_4]}X_1+X_6\\
					X_4+X_5\ce{->[k_5]}X_2+X_6\\
					X_6\ce{->[k_6]}X_5
				\end{array}\]
			\end{minipage}
		}
		\vspace{1.5cm}
		\caption{}
		\label{fig:HK_network}
	\end{subfigure}
	\hfill
	\begin{subfigure}[b]{0.32\textwidth}
		\centering
		\resizebox{\linewidth}{!}{
			\begin{minipage}{\linewidth}
				\begin{align*}
					k_4x_3x_5-k_1x_1 &= 0\\
					k_5x_4x_5+k_1x_1-k_2x_2 &= 0\\
					-k_4x_3x_5+k_2x_2-k_3x_3 &= 0\\
					-k_4x_3x_5-k_5x_4x_5+k_6x_6 &= 0\\
					x_1+x_2+x_3+x_4-k_7 &= 0\\
					x_5+x_6-k_8 &= 0	
				\end{align*}
			\end{minipage}
		}
		\vspace{1cm}
		\caption{}
		\label{fig:HK_system}
	\end{subfigure}
	\hfill
	\begin{subfigure}[b]{0.32\textwidth}
		\includegraphics[width=0.95\linewidth]{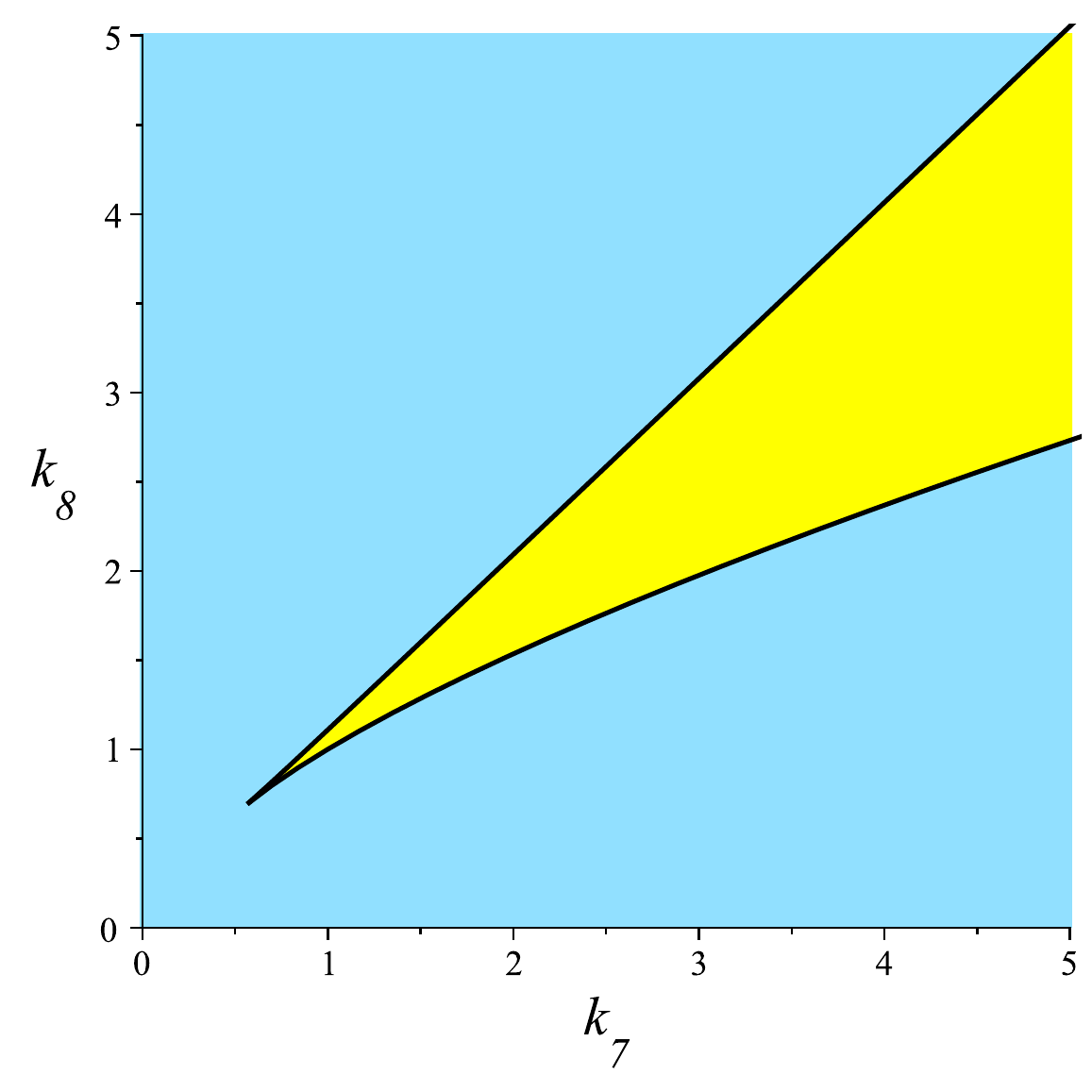}
		\caption{}
		\label{fig:HK_1v2p_CAD}
	\end{subfigure}
	\vskip\baselineskip
	\begin{subfigure}[b]{0.32\textwidth}
		\includegraphics[width=\linewidth]{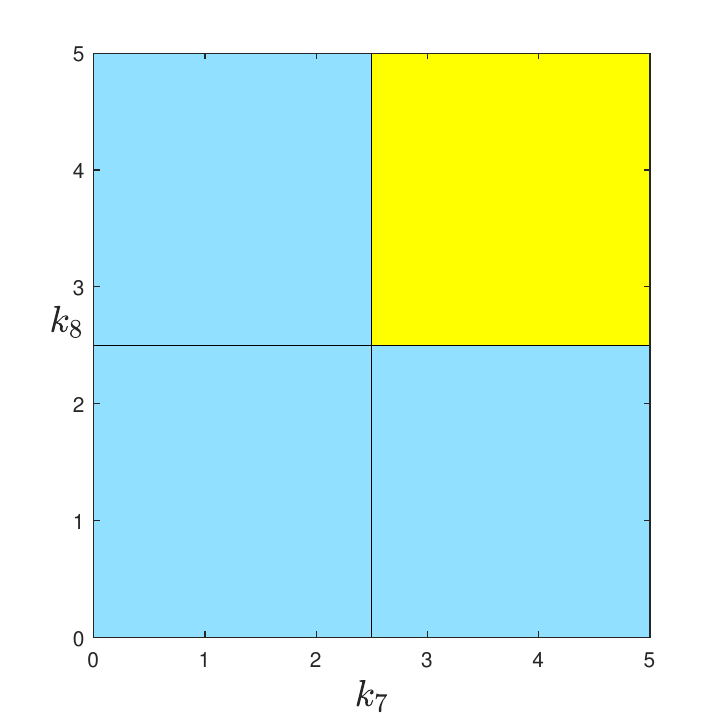}
		\caption{}
		\label{fig:HK_1v2p_Bisect_2valued_plot_n3}
	\end{subfigure}
	\hfill
	\begin{subfigure}[b]{0.32\textwidth}
		\includegraphics[width=\linewidth]{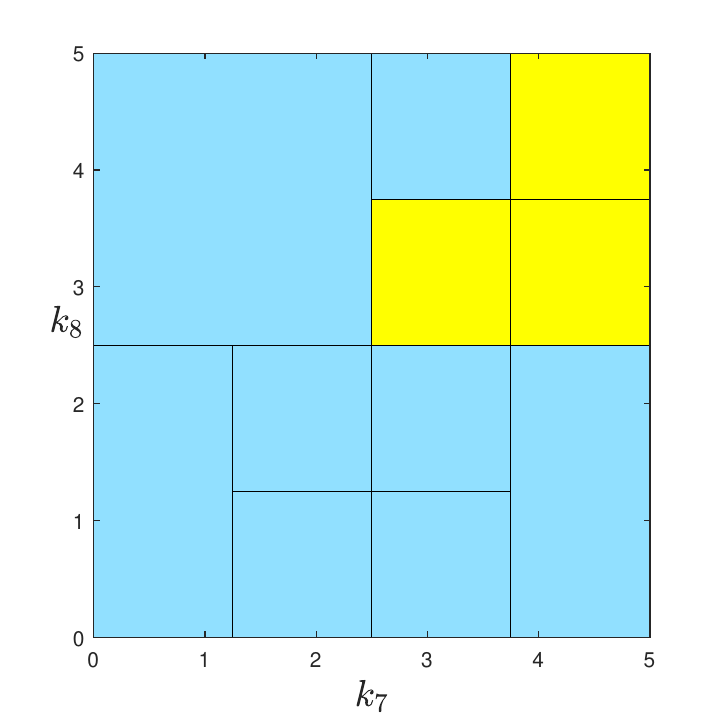}
		\caption{}
		\label{fig:HK_1v2p_Bisect_2valued_plot_n4}
	\end{subfigure}
	\hfill
	\begin{subfigure}[b]{0.32\textwidth}
		\includegraphics[width=\linewidth]{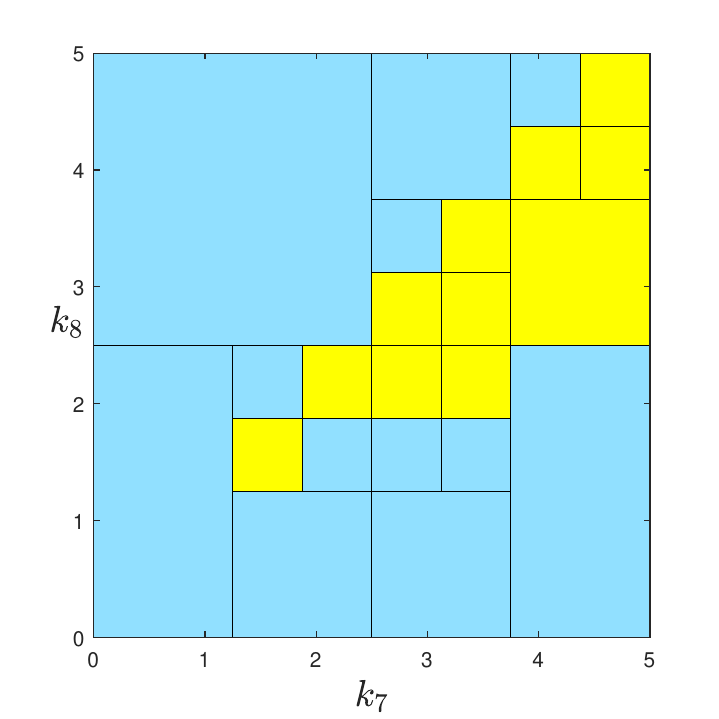}
		\caption{}
		\label{fig:HK_1v2p_Bisect_2valued_plot_n5}
	\end{subfigure}
	\vskip\baselineskip
	\begin{subfigure}[b]{0.32\textwidth}
		\includegraphics[width=\linewidth]{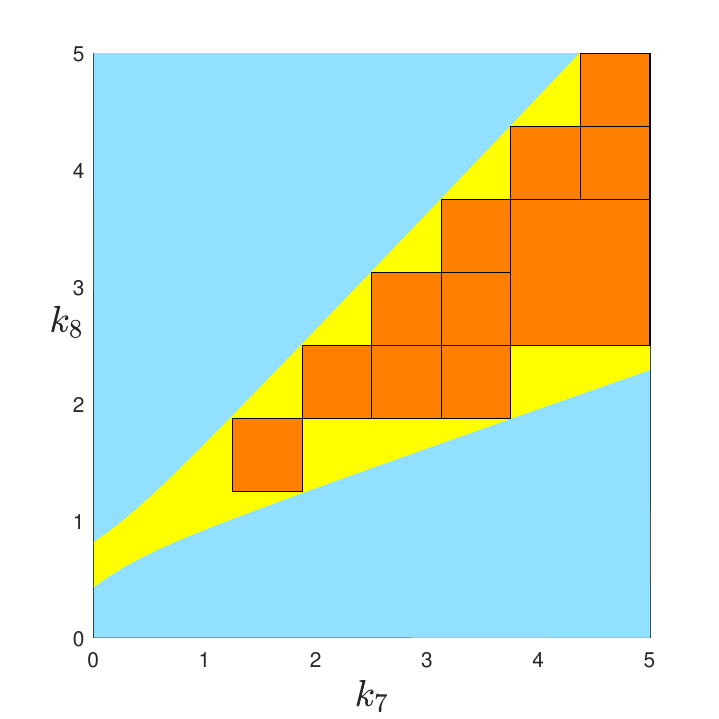}
		\caption{}
		\label{fig:HK_1v2p_PSS_deg_2_from_bisect_rectangular}
	\end{subfigure}
	\hfill
	\begin{subfigure}[b]{0.32\textwidth}
		\includegraphics[width=\linewidth]{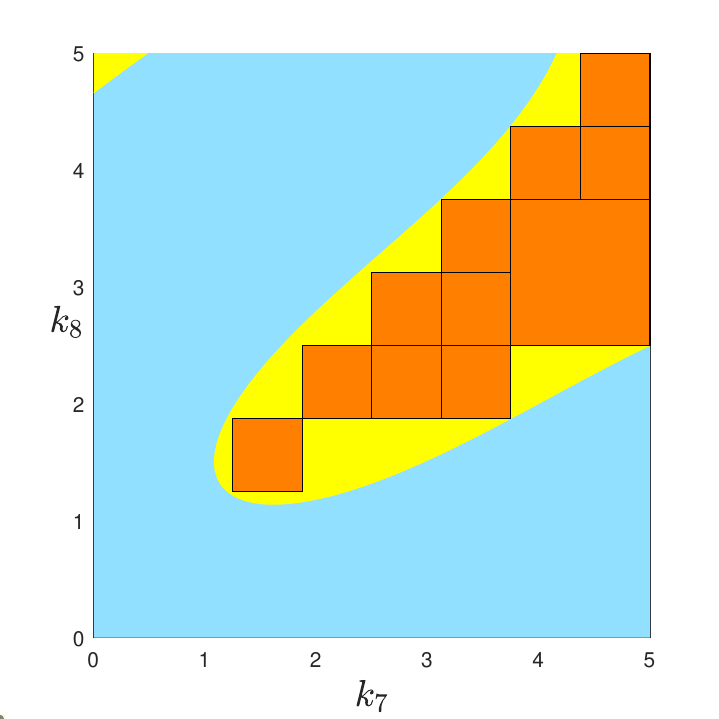}
		\caption{}
		\label{fig:HK_1v2p_PSS_deg_4_from_bisect_rectangular}
	\end{subfigure}
	\hfill
	\begin{subfigure}[b]{0.32\textwidth}
		\includegraphics[width=\linewidth]{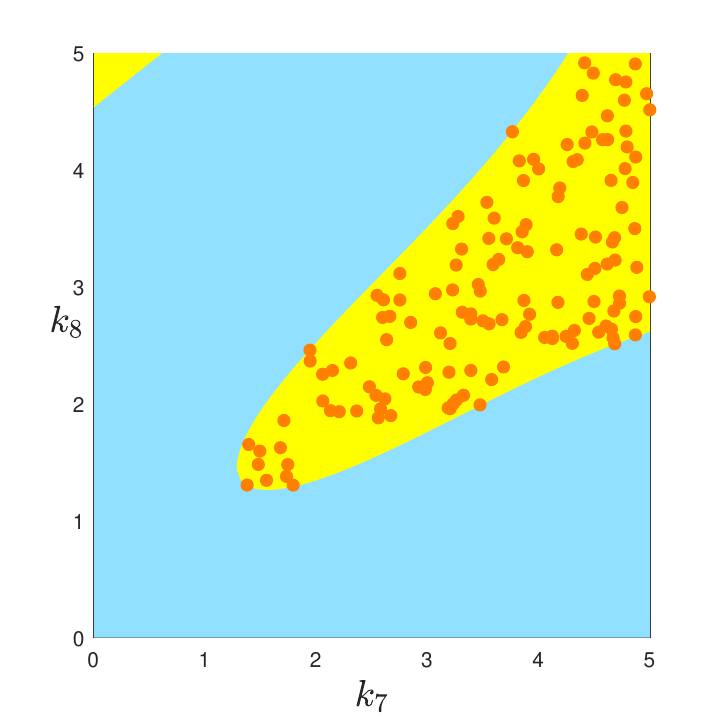}
		\caption{}
		\label{fig:HK_1v2p_PSS_deg_4_from_bisect_sampling}
	\end{subfigure}
	\caption{Using two-valued bisection search to get the PSS representation of the multistationarity region. \\
		(a) The reaction network under study. This network has two conservation laws, thus two of the six steady state equations may be replaced by linear invariants. 
		(b) The system of equations for studying multistationarity. \\
		(c) The CAD representation of the multistationarity region. \\
		(d$-$f) Approximations of the multistationarity region computed by Algorithm \ref{Algorithm:bisect-search-two-value}. \\
		(g$-$h) PSS representations of degrees 2 and 4 computed from the union of yellow rectangles in (f) as K. \\
		(i) PSS representations of degree 4 computed from 140 sample points from the union of yellow rectangles in (f) as approximation of K.
	}
	\label{fig:HK}
\end{figure*}

To illustrate this method we use Example 2.1 of \cite{Feliu-Sadeghimanesh-2020}, shown here in Figure~\ref{fig:HK_network}, for which the Kac-Rice integral is already derived.  The system of equations for studying multistationarity is given in Figure~\ref{fig:HK_system}. Fixing all values of parameters other than $k_7$ and $k_8$ to the following values (similar to the values in \cite{Feliu-Sadeghimanesh-2020}), the goal is to find the multistationarity region of the network in the rectangle $[(0,0),(5,5)]$: 
\begin{equation*}
	k_1=0.7329,\;k_2=100,\;k_3=73.29,\;k_4=50,\;%\\
	k_5=100,\;k_6=5.
\end{equation*}
Figure \ref{fig:HK_1v2p_CAD} shows the exact region computed by CAD. Using Algorithm \ref{Algorithm:bisect-search-two-value} we get the approximation of this region represented in Figures \ref{fig:HK_1v2p_Bisect_2valued_plot_n3}$-$\ref{fig:HK_1v2p_Bisect_2valued_plot_n5}. We see that by decreasing the $\epsilon$ of the termination condition, the approximation is improved. Furthermore using the Kac-Rice integral given in \cite{Feliu-Sadeghimanesh-2020} it takes 1.22, 3.64 and 7.96 seconds for our code written in Julia to compute the approximations in Figures~\ref{fig:HK_1v2p_Bisect_2valued_plot_n3}, \ref{fig:HK_1v2p_Bisect_2valued_plot_n4} and \ref{fig:HK_1v2p_Bisect_2valued_plot_n5} respectively, while solving the system in 1000 points to get the sampling\footnote{To get the rectangular representation by 100 equal subrectangles and solving for 10 points in each subrectangle, it is again necessary to solve the system for 1000 points.} representation in Matlab takes 17.63 seconds. We used Julia for the Kac-Rice integral because it is suggested by \cite{Feliu-Sadeghimanesh-2020} as the fastest platform for this computation. Figure~\ref{fig:HK_1v2p_PSS_deg_2_from_bisect_rectangular} and \ref{fig:HK_1v2p_PSS_deg_4_from_bisect_rectangular} show the PSS approximation of degrees 2 and 4 achieved from Figure \ref{fig:HK_1v2p_Bisect_2valued_plot_n5}. One can also generate random points from the rectangular representation and find the PSS representation from this sampling approximation. Adding the times for using the Kac-Rice integral, two-valued bisection search, generating random points, and computing PSS representation; all together for this example it took 8.53 seconds which is less than finding a rectangular representation by solving the system at 140 points. That result is shown in Figure \ref{fig:HK_1v2p_PSS_deg_4_from_bisect_sampling}.

\begin{remark}\label{Remark:Speed_Bisect_Grid}
	Note that having fewer hyperrectangles in the rectangular representation obtained by the two-valued bisection search does not guarantee a faster speed than the simple approach for finding a rectangular representation discussed in the previous section. Consider the setting in Algorithm~\ref{Algorithm:bisect-search-two-value}. Assume length of all edges of $B\subseteq\R^r$ are the same and equal to $2^m$. Let $\epsilon=1$ and assume $\mathbb{E}\big(g(k)\mid k\sim U(S)\big)$ is not getting close enough to $n_1$ or $n_2$ for any $S$ in the process of this algorithm. Then the total number of expectations that are needed to be computed until the termination of this algorithm is equal to $\sum_{i=0}^{mr}2^i$. On the other hand in the simple approach, by dividing $B$ along each axis to $2^m$ equal parts, the number of needed expectations to be computed is $2^{mr}$.
\end{remark}

\subsection{More general setting}

Now consider a more general case where $\Image(g)=\{n_1,\dots,n_s\}\subseteq\Z_{\geq 0}$. In this case we can not judge about $S\cap L_{n_i}(g)$ just by looking at $\mathbb{E}\big(g(k)\mid k\sim U(S)\big)$. For example if $\Image(g)=\{1,3,5\}$ and we receive $\mathbb{E}\big(g(k)\mid k\sim U(S)\big)=3$, it is not clear that $S$ is an almost subset of $L_3(g)$ or if almost half of it is inside $L_1(g)$ and the other half in $L_5(g)$. So now the goal is to find a way to decide when to add $S$ to $L_i$ when $\mathbb{E}\big(g(k)\mid k\sim U(S)\big)=i$ and when to not and instead bisect it into two sub-hyperrectangles, as in Algorithm \ref{Algorithm:bisect-search-two-value}.

Note that 
\begin{equation*}
	\mathbb{E}\big(g(k)\mid k\sim U(S)\big)=n_1\overline{\Vol}\big(S\cap L_{n_1}(g)\big)+\dots+
	n_s\overline{\Vol}\big(S\cap L_{n_s}(g)\big).
\end{equation*}
Assume $\{n_{\alpha_1},\dots,n_{\alpha_t}\}\subseteq\{n_1,\dots,n_s\}$ such that $\overline{\Vol}\big(S\cap L_{i}(g)\big)\neq 0$ only for $i\in\{n_{\alpha_1},\dots,n_{\alpha_t}\}$. In that case for any distribution on $S$ which has the same zero measure sets as Lebesgue measure's we have $\int_{S\cap L_{i}(g)}q(x)dx=0$ if and only if $i\not\in\{n_{\alpha_1},\dots,n_{\alpha_t}\}$.

Returning to our goal assume $\mathbb{E}\big(g(k)\mid k\sim U(S)\big)=n_i$ for some $i\in\{1,\dots,s\}$. If for every $j\neq i$, $\overline{\Vol}\big(S\cap L_{n_j}(g)\big)=0$, then for any other distribution $q$ on $S$ we have $\int_{S\cap L_{j}(g)}q(x)dx=\delta_{i,j}$, where $\delta_{i,j}$ is 1 if $i=j$, and 0 for $j\neq i$. Therefore $\mathbb{E}\big(g(k)\mid k\sim q\big)=n_i$. Now again assume that $\{n_{\alpha_1},\dots,n_{\alpha_t}\}\subseteq\{n_1,\dots,n_s\}$ such that $\overline{\Vol}\big(S\cap L_{j}(g)\big)\neq 0$ only for $j\in\{n_{\alpha_1},\dots,n_{\alpha_t}\}$. This time let $t\geq 2$. Define the following two sets:
\begin{align*}
	T_1 &= \{(x_1,\dots,x_t)\in (0,1)^t\mid x_1+\dots+x_t=1\},\\
	T_2 &= \{(x_1,\dots,x_t)\in (0,1)^t\mid x_1+\dots+x_t=1,
	n_{\alpha_1}x_1+\dots+n_{\alpha_t}x_t=n_i\}.
\end{align*}
Note that $T_2$ is a set of one dimension lower than dimension of $T_1$. By varying $q$ one can attain any point in $T_1$ by $\big(\int_{S\cap L_{n_1}(g)}q(x)dx,\dots,\int_{S\cap L_{n_t}(g)}q(x)dx\big)$ and $\mathbb{E}\big(g(k)\mid k\sim q\big)=n_i$ if and only if this point belongs to $T_2$. Therefore by probability one for randomly chosen distribution $q$, we will not get $\mathbb{E}\big(g(k)\mid k\sim q\big)=n_i$. Hence we proved the following lemma.

\begin{lem}\label{lemma:Probability_One_lemma}
	Let $B\subseteq\R^r$ be a hyperrectangle and $g\colon B\rightarrow\{n_1,\dots,n_s\}\subseteq\Z_{\geq 0}$. Assume that $\mathbb{E}\big(g(k)\mid k\sim U(B)\big)=n_i$ for some $i\in\{1,\dots,s\}$. Then with probability one we have that $B$ is almost subset of $L_{n_i}(g)$ if and only if $\mathbb{E}\big(g(k)\mid k\sim q\big)=n_i$ for a randomly chosen distribution $q$ on $B$ with the same zero measure sets as Lebesgue measure's.
\end{lem}

Note that in \cite{Feliu-Sadeghimanesh-2020} it is mentioned that the Kac-Rice integral can also be used to compute the expected number of steady states when the parameters are equipped by normal distributions. Even without the Kac-Rice integral, one can solve the system of equations for random sample parameter points chosen from a (truncated) normal distribution and then take the average of the number of solutions. Thus we get Algorithm~\ref{Algoorithm:bisect-search-two-step} which we call \emph{two-step bisection search}.

\IncMargin{1em}
\begin{algorithm}[ht]
	\SetKwInOut{Input}{Input}
	\SetKwInOut{Output}{Output}
	\Input{$B=[a,b]\subseteq\R^r$, $g\colon B\rightarrow\{n_1,\dots,n_s\}\subseteq\Z_{\geq 0}$, $n_1\lneqq\dots\lneqq n_s$, $\epsilon\in\R_{>0}$.}
	\Output{$L_{n_1}(g)\simeq\cup_{i=1}^{m_1}S_{n_1,i}$, $\dots$, $L_{n_s}(g)\simeq\cup_{i=1}^{m_s}S_{n_s,i}$, $\forall i,j\colon S_{n_i,j}\in\mathcal{B}$ and the minimum of the length of the edges of $S_{n_i,j}>\epsilon$.}
	\BlankLine
	\emph{initialization}:
	$L_1=\{\}$, $\dots$, $L_s=\{\}$, $T=\{(B,1)\}$\;
	\While{$T\neq\emptyset$}{choose the first element of $T$ and denote its first element by $S$ and its second element by $i$. Remove $(S,i)$ from $T$\;
		compute $\alpha=\mathbb{E}\big(g(k)\mid k\sim U(S)\big)$\;
		define $\beta$ to be the minimum of the length of the edges of $S$\; 
		\eIf{$\beta\leq\epsilon$}{add $S$ to $L_j$ if $\alpha$ is closer to $n_j$\;}{
			\eIf{$\alpha\not\in\{n_1,\dots,n_s\}$}{define $S_1$ and $S_2$ by dividing $S$ along the $i$-th axis to two equal sub-hyperrectangles. Replace $i$ with $(i+1)\mod n$. Add $(S_1,i)$ and $(S_2,i)$ to $T$\;}{
				choose a distribution on $S$ randomly and call it $q$, for example choose a random point from $S$ and call it $k^\star$, then let $q$ be the truncated normal distribution on $S$ with mean being $k^\star$ and let the variance to be a number not too small and not too large comparing to the length of the edges of $S$\;
				compute $\gamma=\mathbb{E}\big(g(k)\mid k\sim q\big)$\;
				\eIf{$\gamma=n_j$}{add $S$ to $L_j$\;}{define $S_1$ and $S_2$ by dividing $S$ along the $i$-th axis to two equal sub-hyperrectangles. Replace $i$ with $(i+1)\mod n$. Add $(S_1,i)$ and $(S_2,i)$ to $T$\;}
			}
		}
	}
	\caption{Bisection algorithm without restriction on the possible number of solutions for the system of equations.}\label{Algoorithm:bisect-search-two-step}
\end{algorithm}\DecMargin{1em}

\FloatBarrier

\subsection{Example}
\label{example:Lemma_5_3}

Consider the following univariate polynomial of degree 5 with two parameters from \cite[Section 2.3]{Feliu-Sadeghimanesh-2020}.
\begin{align*}
	f(x)=&x^5-(k_1+\tfrac{9}{2})x^4+(\tfrac{9}{2}k_1+\tfrac{21}{4})x^3+(-\tfrac{23}{4}k_1+\tfrac{3}{8})x^2
	\\
	&+(\tfrac{15}{8}k_1-\tfrac{23}{8})x+(\tfrac{1}{100}k_2-\tfrac{1}{16})
\end{align*}
The equation $f(x)=0$ can obtain any number of positive real solutions between 0 and 5 depending on the choice of the parameters. Using the MCKR application \cite{MCKR} which computes the Kac-Rice integral to give the expected number of solutions of a parametric system when the parameters are equipped with a random distribution, we have that the average number of positive solutions of $f(x)=0$ on the two following rectangles is 2 with at least one decimal place accuracy:
\[B_1=[(0.5,8), (1,9)],\quad B_2=[(2,2), (2.5,3)].\]
However, only the first rectangle is inside the parameter region where the number of positive solutions to the system is invariant and equal to 2. Using MCKR, this time we compute the expected number of positive solutions of $f(x)=0$ on $B_2$ when the parameters are equipped with the truncated normal distribution with mean $\mu=(2.25, 2.5)$ and variance $\sigma^2=0.1$. The result with one digit accuracy after the decimal point is 2.9. Thus by Lemma~\ref{lemma:Probability_One_lemma} we can infer the fact that the number of solutions of the system is not invariant in $B_2$ and this set has a non-zero measure subset where the system has something other than 2 positive solutions.

\subsection{Data Access Statement}

All the code and data underpinning the results of this paper is openly available from this URL: https://doi.org/10.5281/zenodo.6927946.

\subsection*{Acknowledgements}

The authors acknowledge the support of EPSRC Grant EP/T015748/1, ``\textit{Pushing Back the Doubly-Exponential Wall of Cylindrical Algebraic Decomposition}". AmirHosein Sadeghimanesh was also supported by NKFIH KKP 129877 when this work was started before moving to Coventry University. Thanks to Hamed Baghal Ghafari for reading and discussing the earlier drafts of the paper and helpful comments about Matlab and \LaTeX.

\end{document}